\documentclass[aps,rmp,reprint,amsmath,amssymb,graphicx,longbibliography]{revtex4-1}

\usepackage{bm}
\usepackage{graphicx}

\begin{document}

\title{Gravitational Waves from Neutrino-Driven Core Collapse Supernovae: Predictions, Detection, and Parameter Estimation}

\author{A. Mezzacappa}
\affiliation{Department of Physics and Astronomy, The University of Tennessee, Knoxville, 1408 Circle Drive, Knoxville, TN 37996 USA}
\author{M. Zanolin} 
\affiliation{Physics and Astronomy Department, Embry-Riddle University, 3700 Willow Creek Road, Prescott, AZ 86301 USA}

\date{\today{}}

\begin{abstract}
\end{abstract}

%\pacs{81.05.Uw,68.37.-d,73.20-r}

\maketitle

\tableofcontents{}

\vspace{-10pt}
\section{Introduction}
\label{sec:Introduction}

In what will hopefully be the not too distant future, Mother Nature will supply us with a Galactic or near-extra-Galactic core collapse supernova (CCSN). This will be the multi-messenger astrophysics event of the century, possibly literally, for which we have all been waiting, and will grace us with {\em all three astrophysical messengers}: gravitational waves (GWs), neutrinos, and photons across the electromagnetic spectrum. The volumes of information that will be brought to us by these cosmic couriers will enable us to validate and, no doubt, greatly improve our models of the supernova central engine, to probe the high-density, neutron-rich physics of the newly forming neutron star, and to probe neutrino physics that may manifest only in such a unique, to say the least, neutrino laboratory.

These are exciting times for both CCSN theory and CCSN observations. For the first time in the nearly sixty years that core collapse supernovae (CCSNe) have been modeled numerically, a consensus across leading CCSN theory groups worldwide is developing: CCSNe can in fact be driven by neutrinos, as originally proposed by Colgate and White \cite{CoWh66} and, then, in the version of the story on which all modern CCSN theory is based, by Bethe and Wilson \cite{BeWi85}. Across numerical methods implemented, CCSN codes that instantiate them, and progenitor characteristics, such as mass, rotation, and metallicity, explosions are now obtained by these groups routinely. The CCSN theory community has entered a new phase of exploration. The challenge now is to obtain results that are in {\em quantitative} agreement with observations (e.g., explosion energies commensurate with observed explosion energies), not {\em qualitative} agreement (e.g., explosion versus the lack of an explosion).

The momentum the worldwide CCSN modeling community now has, and the promise of a multimessenger detection, which could happen at any time, now provide the impetus to the modeling community to address the significant challenges that remain to develop models sufficiently accurate to be able to predict successfully the outcomes {\em for a given progenitor} and, in particular, the progenitor that will be associated with the next Galactic or near-extra-Galactic event. This remains a tall order, and one that relies not only on the CCSN modeling community to fill. CCSN theory is still limited by the lack of three-dimensional progenitor models, although important progress has been made on at least the late stages of stellar evolution. And uncertainties in neutrino weak interaction physics, both theoretical and experimental, and in the high-density, neutron-rich, hot nuclear equation of state must be addressed by others. Nonetheless, decades of progress toward realizing the goal of understanding the CCSN explosion mechanism has taught us one thing: Progress is made, sometimes in increments, sometimes in leaps, by improving the models {\em in each sector} of the macroscopic physics (e.g., the models of gravity, hydrodynamics, magnetohydrodynamics, and neutrino kinetics (transport) implemented) and the microscopic physics (e.g., the neutrino weak interaction rates and nuclear equation of state), one by one. Certainly, the development of high-fidelity models of the macroscopic physics together with the implementation of the state of the art in weak interaction and equation of state physics is the CCSN modeling community's job.

To this end, and duly noting the significant progress that has been made and all of the hard work that went into accomplishing it, it is important to note that only recently have we seen the first published general relativistic CCSN model with state-of-the-art microphysics (see Section \ref{sec:Model Status}). And to be clear, by general relativistic we mean the implementation of general relativistic hydrodynamics or magnetohydrodynamics, general relativistic gravity, and general relativistic neutrino kinetics, and their coupling. As we will discuss in Section \ref{sec:Model Status}, some of the leading models are general relativistic but lack critical microphysics, while others that include the microphysical state of the art implement approximations to general relativity across the major sectors of gravity, hydrodynamics, and neutrino kinetics. Overcoming the shortcomings on either side of this modeling frontier, thereby bridging the two, is the primary challenge CCSN modelers face now. But progress continues to be rapid, and we should have every expectation this gap will be bridged in the not too distant future.

Even with this gap bridged, modelers must face some longer-term challenges that may effect CCSN model outcomes and, with that, our ability to derive the maximal benefit from a Galactic or near-extra-Galactic CCSN detection. In order to get these benefits we also need to have the best possible instruments to record the electromagnetic (EM) emissions, GWs, and neutrinos from these extraordinary events. 

The detection of the first GWs from binary-black-hole mergers in 2015 enabled new comparisons between theoretical predictions and observations. The benefits from similar comparisons in the case of  CCSNe would be enormous, even with the observation of a single CCSN. While other reviews already discuss the value of CCSN EM and neutrino observations, here we focus on the idiosyncrasies of the detection and parameter estimation (PE) from the gravitational wave (GW) channel, using ground-based laser interferometers. Several valuable resources describe the detection and PE software for compact-binary coalescences (CBCs), where the key tool is cross correlation between analytical or numerical deterministic templates and laser interferometric data. The development of algorithms for the detection and PE of CCSN signals on the other hand is more complex both because of the stochastic aspects of the waveforms, as well as the presence of, as we will see, what can only be termed emergent parameters. It is important to notice that what was initially perceived as only a stochastic GW time series has revealed deterministic features both in the time domain and in the time--frequency domain.
Regarding the algorithms for detection and PE, the landscape of methods of which they are composed, both developed and under development, is vast and still growing: likelihood methods, wavelet decomposition, machine learning (ML), and frequentist and Bayesian approaches are some of the more common methods finding application in this setting. We review the current status of the field, including a discussion of current research directions where hypothetical EM and neutrino observations are paired with GW observations in multi-messenger strategies. 

\vspace{-15pt}
\section{The Physics of Neutrino-Driven Core Collapse Supernovae}
\label{sec:Physics}

CCSNe mark the death of massive stars ($M>8-10$ M$_{\odot}$). The iron core of such stars is supported primarily by degenerate relativistic electron pressure, for which a maximum mass, the Chandrasekhar mass, can be supported. As fusion continues in the silicon layer above the iron core, the iron core mass comes to exceed the Chandrasekhar mass, and the core collapses, initiating the supernova. Sound speed decreases with density away from the center and infall velocity increases with distance away from the center. The formation of a sonic point is inevitable, which will set the stage for the formation of the supernova shock wave that will ultimately be responsible for the demise of the progenitor star. The inner, subsonic core proceeds through nuclear matter densities, between $1-2\times 10^{14}$ gcm$^{-3}$, as it collapses and undergoes a phase transition from nucleons and nuclei to nuclear pasta to bulk nuclear matter. At sufficiently high densities, the hard core in the nucleon--nucleon interaction potential, which renders the nucleon--nucleon interaction repulsive, and the Pauli exclusion principle, which also repels like nucleons, lead to a ``stiffening'' of the nuclear equation of state (the pressure increases more rapidly with an increase in density). This stiffening halts the collapse and, moreso, leads to a rebound of the inner core. At the sonic point, information carried by sound about this rebound can no longer propagate outward, and a discontinuity in the flow, with outflow meeting inflow, forms -- i.e., the supernova shock wave.

The so-called ``prompt mechanism'' for CCSNe, where the newly formed shock wave propagates outward, as a dynamical shock, through the outer iron core and the stellar layers above it, reversing the infall of the stellar matter and generating, instead, an outflow, fails. There are two culprits: dissociation and neutrino losses. As the shock propagates through the outer iron core, the nuclei are dissociated by shock compression and heating, costing the shock energy to the tune of $1\times 10^{51}$ ergs per $0.1 M_{\odot}$ of mass dissociated. Then, as the shock passes the electron neutrinosphere, the sphere of last scattering of the electron neutrinos produced through electron capture on nuclei and nucleons during collapse, above which the electron neutrinos are free to escape, carrying energy with them, the shock loses sufficient energy, loses its status as a dynamical shock, and ``stalls,'' becoming instead a standing accretion shock.

The stage is then set for the ``delayed shock mechanism,'' whereby the shock is reenergized to become a dynamical shock. The core becomes stratified. The region below the neutrinospheres (at this point, all three flavors of neutrinos are present) constitutes the proto-neutron star (PNS). Above it, in the cavity between the neutrinospheres and the shock, the matter is heated and cooled by charged-current electron-neutrino interactions involving the nucleons in this cavity, released from nuclei during dissociation. Cooling dominates at higher densities at the base of the cavity, and heating dominates above it, in the region just below the accretion shock. A cooling layer and a heating layer are formed. Neutrino energy deposition in the heating layer is responsible for the revival of the shock.

At core bounce and shock formation and its initial propagation through the core, entropy and lepton gradients are formed, which in turn drive a period of ``prompt (Ledoux) convection'' in the PNS. Heating of the region below the supernova shock wave by neutrinos emerging from the PNS drives ``neutrino-driven (Schwarzchild) convection'' in the heating region. These instabilities assist the neutrinos in generating an explosion by boosting the neutrino luminosities at the neutrinospheres and by boosting the neutrino heating efficiency of the material below the shock. They also generate GWs. These ``local'' instabilities are joined by a ``global'' instability of the supernova shock wave, the so-called Standing Accretion Shock Instability (SASI),'' the latter of which assists neutrino heating by increasing the size of the heating region. The SASI, too, generates GWs. Thus, the very ingredients that lead to explosion by neutrino heating, each have gravitational wave (GW) fingerprints. This will be critical in determining whether or not a CCSN has been detected in GWs, as well as necessary in order to use such a detection to validate CCSN models.

\vspace{-10pt}
\section{The Status of Core Collapse Supernova Models}
\label{sec:Model Status}

Fortunately, we are now in possession of dozens of three-dimensional CCSN models that cut across progenitor characteristics of mass, rotation, and metallicity, and that have implemented different nuclear equations of state. All of these models are quite sophisticated, though all possess major weaknesses. These weaknesses stem from three shortcomings: (1) The use of an effective gravitational potential that is intended to capture the stronger gravitational fields expected from general relativity in an otherwise Newtonian treatment. (2) The use of the so-called ``ray-by-ray'' approximation for neutrino kinetics. (3) The lack of a complete (or nearly complete) set of neutrino interaction physics. By complete, we mean the inclusion of all weak interactions known to be important, or at least nonnegligible, and the implementation of state of the art treatments of them.

The use of an effective potential in the context of CCSNe was first proposed by \citet{RaJa02}. In this approximation, Newtonian hydrodynamics is deployed and the Newtonian gravitational potential is modified. Specifically, the monopole term is replaced by a spherically symmetric potential derived by comparing the Newtonian equation for hydrostatic equilibrium with the Tolman--Oppenheimer--Volkoff equation for general relativistic hydrostatic equilibrium. In a multidimensional context, input to the general relativistic potential is obtained by spherical averaging of the needed quantities, such as the rest mass density, internal energy density, pressure, etc. Liebendoerfer et al. \cite{LiRaJa05} pointed out that the use of an effective potential, at least as originally formulated, overestimates general relativistic effects. As a result, a number of modifications to the procedure were developed and tested (Cases A through N in \citet{MaDiJa06}), with one option, Case A, considered to reproduce best the results of general relativistic simulations. However, a later study by Mueller, Janka, and Marek \cite{MuJaMa12}, in the context of two-dimensional CCSN simulations beginning with the same 15 $M_{\odot}$ progenitor, yielded qualitatively different outcomes for the simulation that was general relativistic and the simulation that relied on the effective potential approach (Case A), with the former yielding an explosion and the latter not. The effective potential approach is used by four separate groups: the Max Planck group, the MSU--Stockholm collaboration, the Princeton group, and the UT--ORNL group. These four groups have produced many of the most sophisticated three-dimensional models published to date. Obviously, this approximation to general relativity has enabled the community to make progress, but no one would argue that we should be satisfied with such an approach and not move toward general relativistic treatments of gravity. Decades of supernova modeling have revealed that model outcomes are sensitive {\em even to small variations} in the input physics or the treatment of the major physics components in the models.

The ray-by-ray approximation to neutrino transport was also first proposed by \citet{RaJa02}. In this approximation, the solution of the three-dimensional neutrino kinetics equations is replaced by the solution of a set of spherically symmetric neutrino kinetics equations, one for each radial ray through each $(\theta,\phi)$ point on the computational grid. This approach was motivated in part by the fact that sophisticated neutrino kinetics solvers had been developed for spherically symmetric simulations of CCSNe. The ray-by-ray approximation is exact, though redundant, for spherically symmetric problems. Thus, the efficacy of the approximation depends on how close to spherical symmetry the problem -- more precisely, the PNS -- remains during the evolution. In axisymmetry, massive accretion funnels persist along the plain perpendicular to the symmetry axis, exciting neutrino emission from the PNS surface in a manner far from spherical symmetry. In these cases, the approximation is not ideal, as discussed by Skinner, Burrows, and Dolence \cite{SkBuDo16}. In three dimensions, without the imposition of a symmetry axis and, therefore, the formation of a small number of massive accretion funnels pinned to equatorial regions of the PNS, the approximation is more effective, as demonstrated by Glas et al. \cite{GlJuJa19}. Two of the above four groups use the ray-by-ray approximation: the Max Planck group and the UT--ORNL group. Again, while the ray-by-ray approximation has enabled important advancements in CCSN theory, particularly in three-dimensional CCSN modeling, no one would argue that we should not move toward three-dimensional solutions to the neutrino kinetics equations.

The history of the deployment of increasingly realistic neutrino weak interaction physics in CCSN models has followed the development of the theory of weak interaction physics itself, especially in the years immediately following the publication of the electroweak theory. In particular, neutral-current weak interactions, which allow coherent scattering of neutrinos on nuclei, fundamentally altered the dynamics of stellar core collapse \cite{Wilson1974}. The later introduction of new weak interactions -- e.g., neutrino pair production by nucleon--nucleon bremsstrahlung \cite{HaRa98} -- opened up new channels for the production of neutrino pairs of all three flavors, and more recent work factoring in nucleon correlations in nuclei \cite{HiMeMe03,LaMaSa03} and nuclear matter \cite{BuSa98,RePrLa98}, all proved to be essential improvements to CCSN models. Thus, a {\em minimal} set of neutrino weak interactions must include (1) electron capture on nuclei, with nucleon correlations taken into account, (2) electron and positron capture on nucleons, including degeneracy and relativity, (3) coherent, isoenergetic scattering on nuclei, (4) nonisoenergetic scattering on nucleons, factoring in the small energy exchange, (5) nonisoenergetic scattering on electrons and positrons, (6) neutrino pair production from electron--positron pairs, and (7) neutrino pair production from nucleon--nucleon bremsstrahlung. Prior to near the turn of the millennium, all models included (1)--(6) but without the effects of nucleon correlations, degeneracy, relativity, and small-energy scattering -- the so-called ``Bruenn85'' opacity set \cite{Bruenn1985}. Some of today's leading three-dimensional models \cite{LeBrHi15,MeJaMa15,BuRaVa19,VaBuRa19}, at least from a microphysical perspective, include the minimal set and, in some cases more (e.g., the inclusion of muons and their associated weak interactions \cite{BoJaLo17}). Until recently, other of today's leading models \cite{RoOtHa16,KuTaKo16}, from a macrophysical perspective [i.e., were three-dimensional and general relativistic (or nearly so)], did not include the minimal set (they include the Bruenn85 set or the Bruenn85 set plus pair production from nucleon--nucleon bremsstrahlung). 

\citet{Kuroda2021} finally advanced the frontier to perform the first three-dimensional, general relativistic CCSN simulations with an extensive set of neutrino interactions.

\vspace{-15pt}
\section{Sources of Core Collapse Supernova Gravitational Waves}
\label{sec:Sources}

At this point in time, a number of studies of CCSN GW emission have been conducted based on simulation data from three-dimensional models 
\cite{
AfKuCa23, 
Andresen_2017, 
Andresen:2018aom, 
BrBiOb23, 
DrAnDi23, 
HaKuKo15, 
HaKuKo18, 
HaKuNa16, 
KaKuTa18, 
KoIwOh11, 
KoIwOh09, 
KuFiTa22, 
KuKoHa17, 
KuKoTa16, 
Kuroda_2018, 
LiRiLu23, 
MaAsTa22, 
MeMaLa20, 
MeMaLa23, 
Nakamura_2022, 
OCCo18, 
Ott_2013, 
PaVaPa23, 
PaWaCo21, 
Pan_2018, 
Pan_2021, 
Powell_2019, 
PoMu20, 
PoMu22, 
RaMoBu19, 
RiZaAn23, 
RiOtAb17, 
Scheidegger_2010, 
ShKuKo20, 
ShKuKo21, 
SrBaBr19, 
TaKo18, 
VaBu20, 
VaBuRa19, 
VaBuWa23, 
Warren_2020}. 
Given the results of these studies, and the conclusions reached, the sources of GWs in CCSNe can be broken down as follows:

\begin{figure}
\includegraphics[width=\columnwidth,clip]{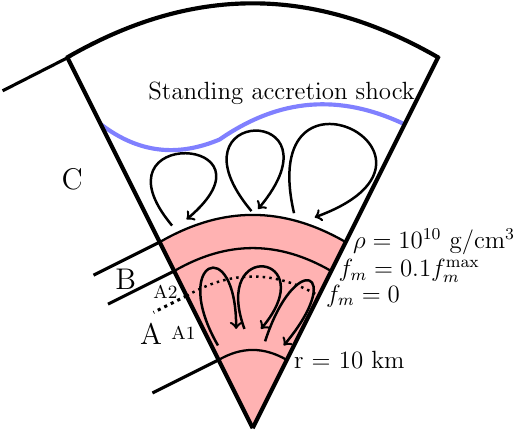}
\caption{Sources of CCSN GW emission (taken from \citet{Andresen_2017}), beginning with sustained Ledoux convection in Region A$_1$ deep within the PNS, moving outward to convective overshoot in Region A$_2$, then on to neutrino-driven convection in the gain layer, Region C, which generates GWs {\em per se} but also leads to accretion plumes impinging on the PNS surface. These plumes in turn excite GW emission from the PNS. Finally, the SASI modulates the flow in Region C, as well as the accretion onto the PNS surface, indicated here by the $\rho=10^{10}$ gcm$^{-3}$ contour, generating low-frequency GW emission.}
\label{fig:gravitationalwavesources}
\end{figure}

{\em Core Bounce}: We expect the majority of CCSN progenitors to be rotating, albeit slowly. In the presence of such rotation, stellar core bounce results in a significant time-dependent quadrupole moment, which in turn produces the first significant feature in the GW strain. This feature comprises a prominent spike in the strain amplitude, together with secondary peaks associated with bounce-induced oscillations of the PNS (see, for example, \citet{RiOtAb17}. In most of the studies cited above, the bounce signature is not present in the waveforms given that rotation was not included in the associated models. In the absence of three-dimensional progenitors, it is difficult to know what the differential core rotation is across the progenitors considered.

{\em Prompt Convection in the Proto-Neutron Star}: This is convection that occurs immediately after bounce, due to the gradients in entropy and lepton fraction left behind by (i) the weakening shock and (ii) the core deleptonization during collapse and the escape of copious electron neutrinos during the electron-neutrino burst within the first $\sim$5 ms after core bounce, respectively. It is short-lived, normally occurring between 5 and 20 ms after bounce. The convection flattens the gradients that give rise to it. Prompt convection gives rise to high-frequency ($>500$ Hz) GW emission.

{\em Sustained Proto-Neutron Star Ledoux Convection}: Unlike prompt convection, PNS Ledoux convection is sustained by lepton gradients that are sustained by the continuing deleptonization of the PNS by neutrino diffusion out of it. It begins at about 100 ms after bounce and is sustained for the duration of the simulations. Like prompt convection, PNS convection gives rise to high-frequency GW emission.
    
{\em Convective Overshoot from Sustained Proto-Neutron Star Ledoux Convection}: This results from the convective overshoot into the convectively stable layer above the Ledoux convection layer. Its contribution to the high-frequency GW emission is typically larger than the emission emanating from the convective region itself. Here too, the GW emission is of high frequency.

{\em Neutrino-Driven Convection in the Gain Layer}: Shock revival is mediated by heating of the material below the stalled shock wave by electron-neutrino and electron-antineutrino absorption on the dissociation-liberated neutrons and protons, respectively. Because of the $r^{-2}$ dependence of the neutrino flux emanating from the PNS surface (the neutrinospheres), the heating is greatest at the base of the heating or ``gain'' layer. As a result, an entropy gradient develops in this layer and Schwarzchild convection follows. The convection becomes turbulent. Turbulent neutrino-driven convection gives rise to low-frequency ($<250$ Hz) GW emission.
    
{\em The Standing Accretion Shock Instability (SASI)}: The stalled supernova shock wave is unstable to non-spherical perturbations. The perturbations grow, leading to ``sloshing'' \cite{BlMeDe03} and ``spiraling'' \cite{BlMe07} motions beneath it, which in turn are reflected in the shape of the shock wave itself as the explosion develops. Like turbulent neutrino-driven convection, the SASI gives rise to low-frequency GW emission.
    
{\em Accretion onto the Proto-Neutron Star Surface}: Given turbulent neutrino-driven convection and the SASI, the accretion onto the PNS, which occurs via accretion plumes that penetrate the gain radius separating the gain layer and the neutrino cooling layer below it, impinging on the PNS surface and exciting high-frequency GW emission.
    
{\em Explosion}: The non-spherical explosion of material gives rise to a low-frequency time-dependent quadrupole moment and, consequently, very-low-frequency ($<50$ Hz) GW emission.

Most of the eight sources listed above are illustrated in Figure \ref{fig:gravitationalwavesources}.

As we will discuss in Section \ref{sec:Predictions}, predictions for high- and low-frequency GW emission across the studies conducted to date by various groups are qualitatively very similar. On the other hand, conclusions regarding the sourcing of this GW emission have varied considerably across the same studies.

\vspace{-10pt}
\section{Calculating Core Collapse Supernova Gravitational Wave Emission: Formalism}
\label{Formalism}

The Einstein equations for the spacetime metric, $g_{\mu\nu}$, in the line element

\begin{equation}
ds^2=g_{\mu\nu}dx^{\mu}dx^{\nu}
\label{eq:line element}
\end{equation}
are given in natural units by 

\begin{equation}
G_{\mu\nu}=8\pi T_{\mu\nu},
\label{eq:einstein equations}
\end{equation}
where the Einstein tensor, $G_{\mu\nu}$, is 

\begin{equation}
G_{\mu\nu}=R_{\mu\nu}-\frac{1}{2}g_{\mu\nu}R,
\label{eq:einstein tensor}
\end{equation}
and the Ricci tensor, $R_{\mu\nu}$, and the Ricci scalar, $R$, are 

\begin{equation}
R_{\mu\nu}=R^{\lambda}_{\mu\lambda\nu}
\label{eq:ricci tensor}
\end{equation}
and

\begin{equation}
R=g^{\mu\nu}R_{\mu\nu}.
\label{eq:ricci scalar}
\end{equation}
The dependence on the metric implicit in the quantities above is made explicit by considering the Riemann curvature tensor, $R^{\lambda}_{\mu\beta\nu}$,

\begin{equation}
R^{\lambda}_{\mu\beta\nu}=
\partial_{\beta}\Gamma^{\lambda}_{\mu\nu}-
\partial_{\nu}\Gamma^{\lambda}_{\mu\beta}+
\Gamma^{\lambda}_{\alpha\beta}\Gamma^{\alpha}_{\mu\nu}-
\Gamma^{\lambda}_{\alpha\beta}\Gamma^{\alpha}_{\mu\beta},
\label{eq:riemann curvature tensor}
\end{equation}
which is expressed in terms of the connection coefficients, $\Gamma^{\lambda}_{\mu\nu}$,

\begin{equation}
\Gamma^{\lambda}_{\mu\nu}=\frac{1}{2}g^{\lambda\rho}\left(
\partial_{\nu}g_{\mu\rho}+
\partial_{\mu}g_{\nu\rho}-
\partial_{\rho}g_{\mu\rho}\right).
\label{eq:christoffel symbols}
\end{equation}

In the case of CCSNe, GWs can be considered as linear perturbations of the background spacetime geometry. Choosing the background geometry to be the geometry far from the source, we write

\begin{equation}
g_{\mu\nu}=\eta_{\mu\nu}+h_{\mu\nu}.
\label{eq:linearized metric}
\end{equation}
Upon substitution of this metric in the Einstein equations, Eqn. (\ref{eq:einstein equations}), we obtain a wave equation for the metric perturbation, $h_{\mu\nu}$,

\begin{equation}
\Box \bar{h}_{\mu\nu}=-16\pi T_{\mu\nu},
\label{eq:wave equation}
\end{equation}
where

\begin{equation}
\bar{h}_{\mu\nu}=h_{\mu\nu}-\frac{1}{2}hg_{\mu\nu},
\label{eq:hbar}
\end{equation}
and where $h$ is the trace of $h_{\mu\nu}$,

\begin{equation}
h=g^{\mu\nu}h_{\mu\nu}.
\label{eq:trace of h}
\end{equation}
In the vacuum case, Eqn. (\ref{eq:wave equation}) becomes (from now on we remove the bar notation)

\begin{equation}
\Box h_{\mu\nu}=0,
\label{eq:wave equation_freespace}
\end{equation}
with solutions of the form 

\begin{equation}
h_{\mu\nu}=e_{\mu\nu}e^{ik\cdot x},
\label{eq:vaccum solution}
\end{equation}
where $e_{\mu\nu}$ is the polarization tensor, which for a GW propagating along the $\hat{\mathbf{e}}_{z}$ direction is

\begin{equation}
e_{\mu\nu}=\left(
\begin{matrix}
h_+ & h_\times & 0 \\
h_\times & -h_+ & 0 \\
0 & 0 & 0
\end{matrix}
\right).
\label{eq:z-axis vacuum solution}
\end{equation}
$h_+$ and $h_\times$ represent the amplitudes of the two polarizations of the GW, the so-called ``plus'' and ``cross'' polarizations, respectively.

In the presence of a source, the solution to Eqn. (\ref{eq:einstein equations}) can be written as

\begin{equation}
\bar{h}_{\mu\nu}=-16\pi\int d^{4}x'G(x-x')T_{\mu\nu}(x'),
\label{eq:general solution}
\end{equation}
where the Green's function, $G(x-x')$, satisfies

\begin{equation}
\Box G(x-x')=\delta^{(4)}(x-x'),
\label{eq:green's function equation}
\end{equation}
whose solution is

\begin{equation}
G(x-x')=-\frac{1}{4\pi |\mathbf{x}-\mathbf{x}'|}\delta(t-|\mathbf{x}-\mathbf{x}'|-t').
\label{eq:green's function}
\end{equation}
Substituting Eqn. (\ref{eq:green's function}) in Eqn. (\ref{eq:general solution}), we obtain

\begin{equation}
\bar{h}_{\mu\nu}(\mathbf{x},t)=4\int d^{3}x' \frac{T_{\mu\nu}(\mathbf{x}',t-|\mathbf{x}-\mathbf{x}'|)}{|\mathbf{x}-\mathbf{x}'|},
\label{eq:final solution}
\end{equation}
where the integration is over the source volume.

We now adopt the slow motion approximation and expand $T_{\mu\nu}$ in Eqn. (\ref{eq:final solution}) in powers of $v/c$. In addition, we perform a transverse traceless decomposition of the metric to arrive at (putting back the factors of $c$)

\begin{equation}
h^{TT}_{ij}=\frac{G}{c^4}\frac{1}{r}\sum_{m=-2}^{+2}\frac{d^{2}I_{2m}}{dt^2}(t-\frac{r}{c})f^{2m}_{ij},
\end{equation}
where

\begin{equation}
I_{2m}=\frac{16\sqrt{3}\pi}{15}\int \rho Y^{*}_{2m}r^{2}dV
\label{eq:quadrupole moment}
\end{equation}
is the mass quadrupole moment and

\begin{equation}
f^{lm}_{ij}=r^{2}\left(
\begin{matrix}
0 & 0 & 0 \\
0 & W_{lm} & X_{lm} \\
0 & X_{lm} & -\sin^{2}\theta\, W_{lm}
\end{matrix}
\right)
\label{eq:tensor spherical harmonics}
\end{equation}
are the tensor spherical harmonics, with

\begin{equation}
X_{lm}=2\frac{\partial}{\partial\phi}\left(\frac{\partial}{\partial\theta}-\cot\theta\right)Y_{lm}(\theta,\phi)
\label{eq:xlm}
\end{equation}
and

\begin{equation}
W_{lm}=\left(\frac{\partial^{2}}{\partial\theta^{2}}-\cot\theta\frac{\partial}{\partial\theta}-\frac{1}{\sin^{2}\theta}\frac{\partial^{2}}{\partial\phi^{2}}\right)Y_{lm}(\theta,\phi).
\label{eq:wlm}
\end{equation}
The gravitational wave ``strain'' amplitudes of the plus and cross polarizations are then given by

\begin{equation}
h_{+}=\frac{h_{\theta\theta}}{r^2}
\label{eq:hplus}
\end{equation}
and

\begin{equation}
h_{\times}=\frac{h_{\theta\phi}}{r^{2}\sin\theta}.
\label{eq:hcross}
\end{equation}
For additional details, we refer the reader to \citet{KoSaTa06}.

\vspace{-10pt}
\section{Predictions of Core Collapse Supernova Gravitational Wave Emission from Multi-Physics Simulations}
\label{sec:Predictions}

\subsection{General Characteristics of Core Collapse Supernova Gravitational Wave Strains}
\label{sec:Predictions_GeneralCharacteristicsStrains}

\begin{figure}
\vspace{-24pt}
\includegraphics[width=0.49\textwidth]{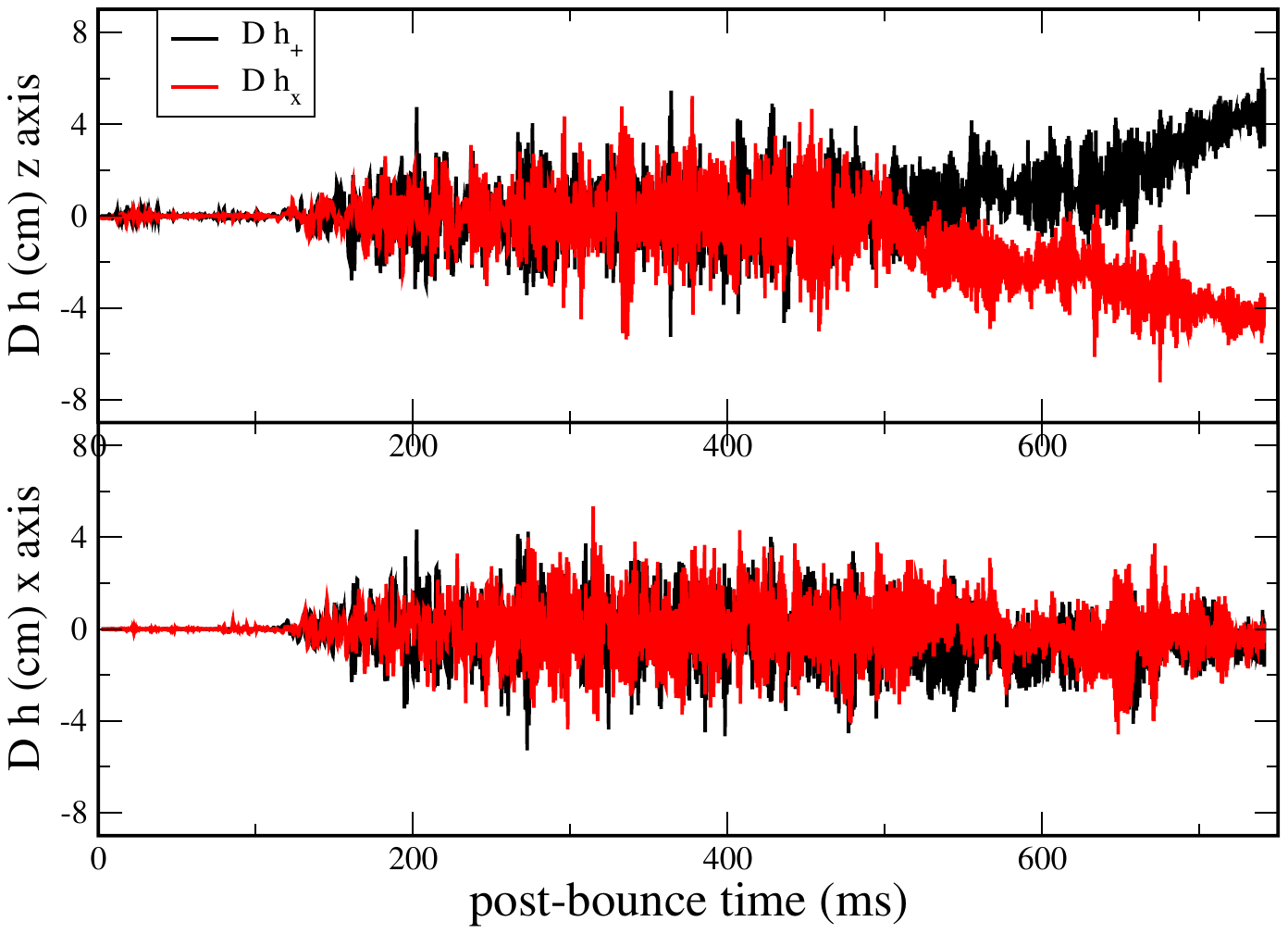}
\vspace{-20pt}
\caption{CCSN GW strains (at the source) for both polarizations for a progenitor of mass 15 M$_\odot$, from \citet{MeMaLa23}.}
\label{fig:Mezzacappaetal2023Strains}
\end{figure}

Figure \ref{fig:Mezzacappaetal2023Strains} shows the plus- and cross-polarization GW strains (at the source) for a 15 M$_\odot$ model from \citet{MeMaLa23}. 
Given the lack of rotation, there is no bounce signal and, with the exception of some low-amplitude emission from prompt convection, GW emission does not begin until $\sim$100 ms after bounce, when several excitation mechanisms set in. By this time, sustained Ledoux convection within the PNS has begun, and neutrino-driven convection and the SASI have developed, all three of which in turn excite high-frequency GW emission from the PNS. This period of significant high-frequency emission is supplemented by an overall offset of the strain after $\sim$500 ms post-bounce. This offset, or ``memory'' results from explosion.

\vspace{-15pt}
\subsection{General Characteristics of Core Collapse Supernova Spectrograms}
\label{sec:Predictions_GeneralCharacteristicsSpectrograms}

Figure \ref{fig:Mezzacappaetal2023Heatmaps} shows the spectrogram for the model considered above. 
Focusing first on the time period $t>100$ ms after bounce, we see that the spectrogram exhibits two primary features: The first, a component above $\sim$500 Hz that rises in frequency with post-bounce time. The second, a component below $\sim$250 Hz that remains within a constant range in frequency with post-bounce time. The high-frequency component of the spectrogram originates in the PNS. It is excited by prompt convection, sustained Ledoux convection, and sustained Ledoux convection overshoot in the PNS, as well as accretion onto the PNS. The low-frequency component can originate from the gain layer due to the SASI's impact on the flow in this region, and/or from low-frequency modulation of the accretion onto the PNS by the SASI. At later times, at very low frequencies, the GW emission originates from explosion itself.

The situation is dramatically different for $t<100$ ms, where little GW emission is seen. This is in part an artifact of the use of non-rotating and non-perturbed progenitors (i.e., spherically symmetric progenitors). \citet{MeMaLa23} document a dramatically different case, initiated from a 9.6 M$\odot$ progenitor. During collapse, this model exhibits nuclear burning as a result of its composition, which perturbs the core prior to bounce, yielding significant GW emission early after bounce. 
This, together with earlier work by \citet{CoChAr15}, \citet{MuMeHe17}, and \citet{VaCoBu22}, highlights the need for three-dimensional progenitor models for both explosion modeling and predicting accurately this early phase of GW emission. Until these models are available, confidence in our predictions of CCSN GW strains and spectrograms will be higher at later times after bounce, where agreement across model predictions by different groups has been demonstrated. 

\vspace{-10pt}
\subsection{Sourcing Gravitational Wave Emission in Core Collapse Supernovae}
\label{sec:sourcing}

\begin{figure}
\includegraphics[width=\columnwidth,clip]{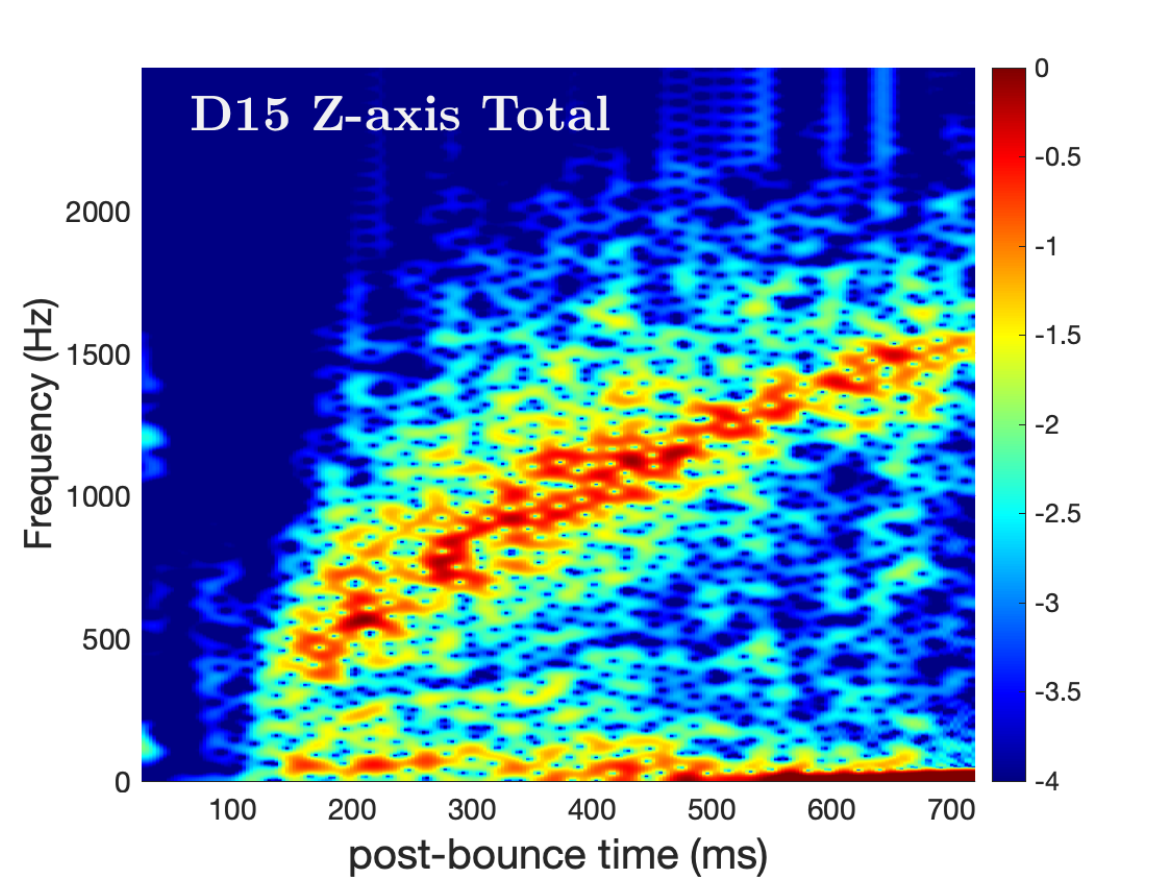}
\caption{Spectrogram for the model considered in Figure \ref{fig:Mezzacappaetal2023Strains}.}
\label{fig:Mezzacappaetal2023Heatmaps}
\end{figure}

\begin{figure}
\includegraphics[width=\columnwidth,clip]{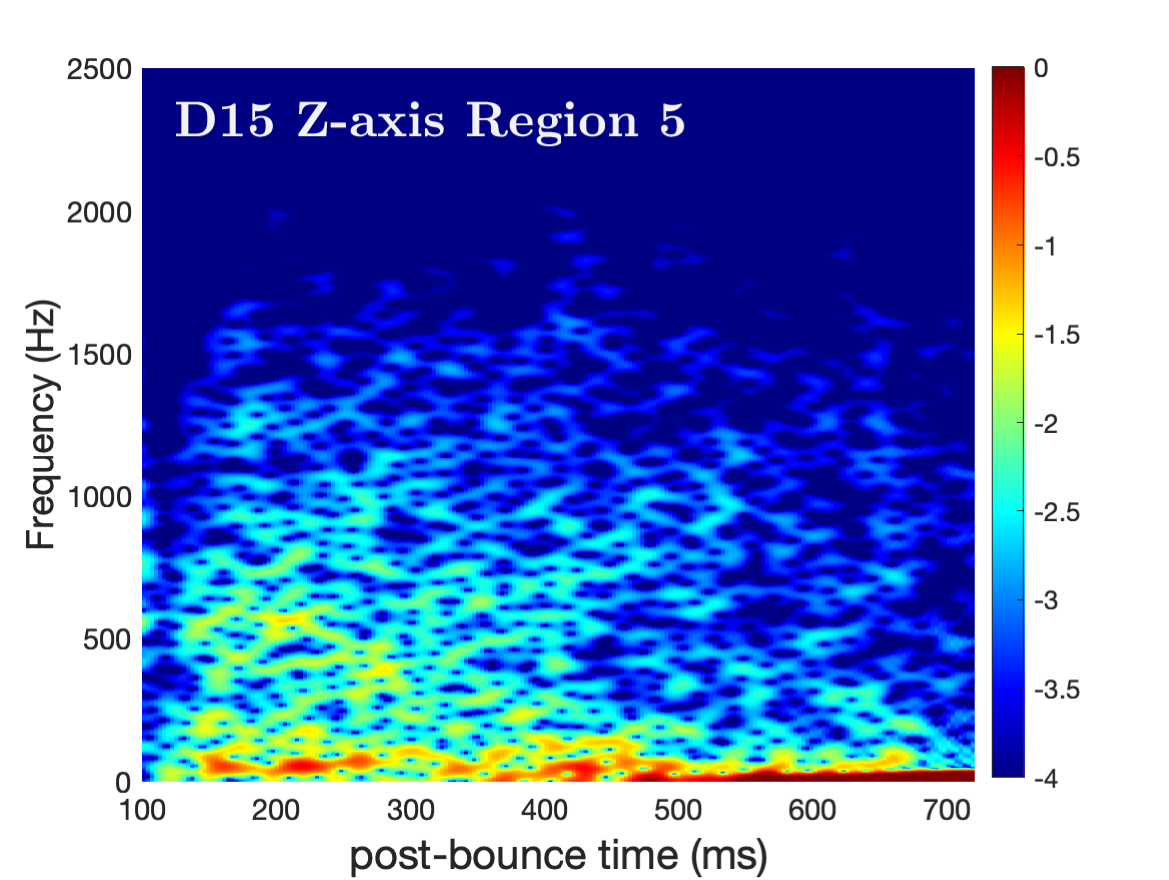}
\caption{Region R5 spectrogram for the model considered in Figure \ref{fig:Mezzacappaetal2023Strains}.}
\label{fig:Mezzacappaetal2023-15-R5-Heatmap}
\end{figure}

\begin{figure}
\includegraphics[width=\columnwidth,clip]{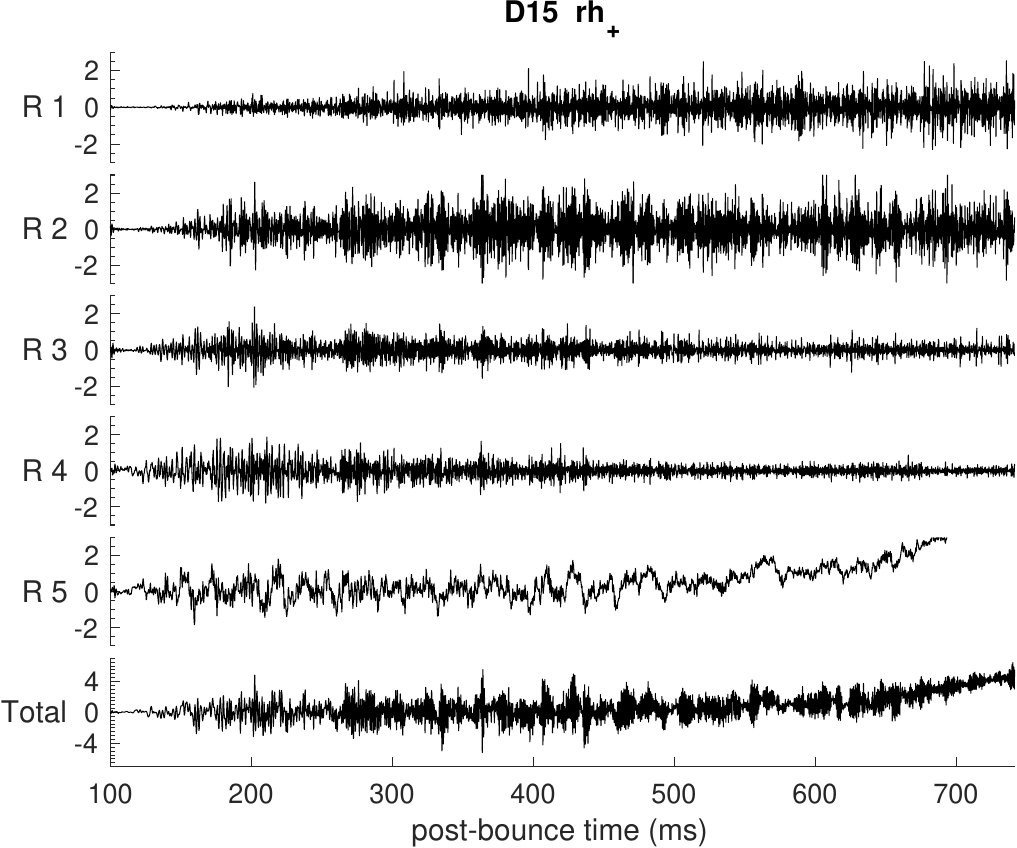}
\caption{GW strain by region for the model considered in Figure \ref{fig:Mezzacappaetal2023Strains}.}
\label{fig:Mezzacappaetal2023RegionalStrains}
\end{figure}

GW emission from neutrino-driven CCSNe is directly linked to the  components of the central engine driving such supernovae. The construction of a detailed GW strain from a detection of the next Galactic or near-extra-Galactic event will then provide an opportunity to probe these components and to validate our models of them. However, this will rely on our ability to source spatially the GW emission from the center of the PNS outward. Thus far, both indirect and direct methods have been developed to do so.

In the context of three-dimensional CCSN models, direct methods to source spatially the GW emission have been developed by \citet{Andresen_2017} and \citet{MeMaLa20,MeMaLa23}. Figure \ref{fig:Mezzacappaetal2023RegionalStrains} shows the strain amplitude for the 15 M$_\odot$ model of \citet{MeMaLa23}. The strain is computed for five distinct regions, R1 through R5, corresponding to A$_1$ (R1), A$_2$ (R2), B (R3), and C (R4 and R5) in Figure \ref{fig:gravitationalwavesources}. The evolution of the high-frequency emission and its sources is made clear by this spatial decomposition. Over the first several hundred milliseconds after bounce, it emanates initially from R4. The emission then moves inward to region R3 -- i.e., the PNS surface layer -- and R2, excited by accretion onto the PNS. However, at later post-bounce times, the high-frequency emission emanates largely from deep within the PNS, from regions R1 and R2 -- i.e., the region of Ledoux convection and the convective overshoot region above it. This one example illustrates how complex and time dependent the emission and sourcing in a model can be. In this model, there is no single source or region that is responsible for the high-frequency GW emission throughout the evolution.

Focusing now on the low-frequency emission in this model, Figure \ref{fig:Mezzacappaetal2023-15-R5-Heatmap} shows the spectrogram for the same 15 M$_\odot$ model but for region R5 only. Comparing Figures \ref{fig:Mezzacappaetal2023Heatmaps} and \ref{fig:Mezzacappaetal2023-15-R5-Heatmap}, it is evident that in this model low-frequency emission stems from region R5, the gain layer. For Andresen et al., on the other hand, the low-frequency emission stemmed from the PNS as the result of the low-frequency modulation of accretion onto the PNS by the SASI.

\begin{figure}
\includegraphics[width=\columnwidth,clip]{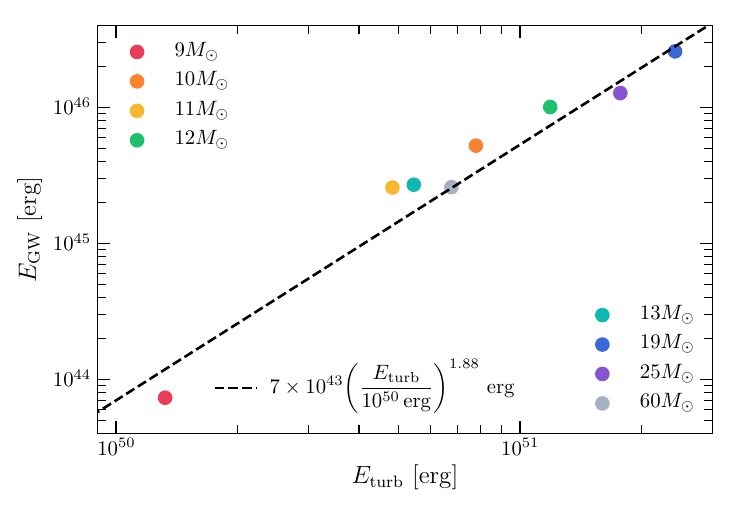}
\caption{Total GW energy released as a function of the energy (kinetic and thermal) accreted by the PNS, for the models considered in Figure 6 of \citet{RaMoBu19}.}
\label{fig:GWvTurbulentEnergy}
\end{figure}

Moving now to indirect methods that have been developed to source CCSN GW emission, we consider the example provided by \citet{RaMoBu19}. Specifically, these authors computed the energy radiated in GWs versus the turbulent energy accreted onto the PNS for four models of 9, 10, 11, and 12 M$_\odot$ they considered. The strong correlation between the two, shown in Figure \ref{fig:GWvTurbulentEnergy}, led \citet{RaMoBu19} to conclude that the GW emission is sourced in these models primarily by accretion onto the PNS by turbulent-convection-- and SASI--induced funnels.

\vspace{-10pt}
\subsection{Culling Information about the Proto-Neutron Star through a Gravitational Wave Detection}
\label{sec:normalmode}
In light of the discussion in the previous section, the detection of GW emission from a CCSN will enable us to observe the fluid instabilities at play that aid the neutrinos in powering an explosion. The hope, too, is that a detection would also enable us to say something about the PNS remnant -- e.g., its mass and radius. For this, the GW astrophysics community has turned to asteroseismology.

Applications of asteroseismology to CCSNe, more specifically the PNS, were first carried out by \citet{FeMiPo03} and later by \citet{FuKlAb15}, Sotani and collaborators \cite{SoTa16,SoKuTa17,SoKuTa19,SoTa20,SoTaTo21}, Torres-Forne' and collaborators \cite{ToCePa18,ToCeOb19,ToCePa19}, Morozova and collaborators \cite{MoRaBu18,RaMoBu19}, Westernacher-Schneider and collaborators \cite{WeOcSu19,Westernacher-Schneider20}, and others (e.g., most recently by \citet{AfKuCa23} and \citet{MoSuTa23}).
This work has led to the association of specific features of the CCSN spectrograms computed by various groups with specific modes: $g$-modes, mediated by gravity, $p$-modes, mediated by pressure, and the fundamental or $f$-mode, also mediated by pressure. Arguably the prominent high-frequency feature observed in all CCSN spectrograms computed to date is a focus of CCSN GW detection and PE, in the latter case particularly with regard to estimating the mass and radius of the PNS.

It is well known from asteroseismology that the peak frequency of $g$-modes depend on the gravity at the source -- i.e., 
\begin{equation}
    f_{\rm peak}\sim \frac{GM}{R^2},
\label{eq:gmodepeak}
\end{equation}
where $G$ is Newton's gravitational constant, $M$ is the enclosed mass, and $R$ is the radius, at which the modes are excited. On the other hand, the peak frequency of an $f$-mode depends on the square root of the mean density of the star -- i.e., 
\begin{equation}
    f_{\rm peak}\sim\sqrt{\frac{M}{R^3}}\,,
\label{eq:fmodepeak}
\end{equation}
where in this case $M$ and $R$ are the stellar mass and radius, respectively. If we are to use a CCSN GW detection -- specifically the peak frequency of the detected GWs and its time evolution -- to determine the mass and/or radius of the PNS and, in turn, to learn something about its equation of state, we must first know what modes are giving rise to the peak frequency emission. 

\begin{figure}
\includegraphics[width=\columnwidth,clip]{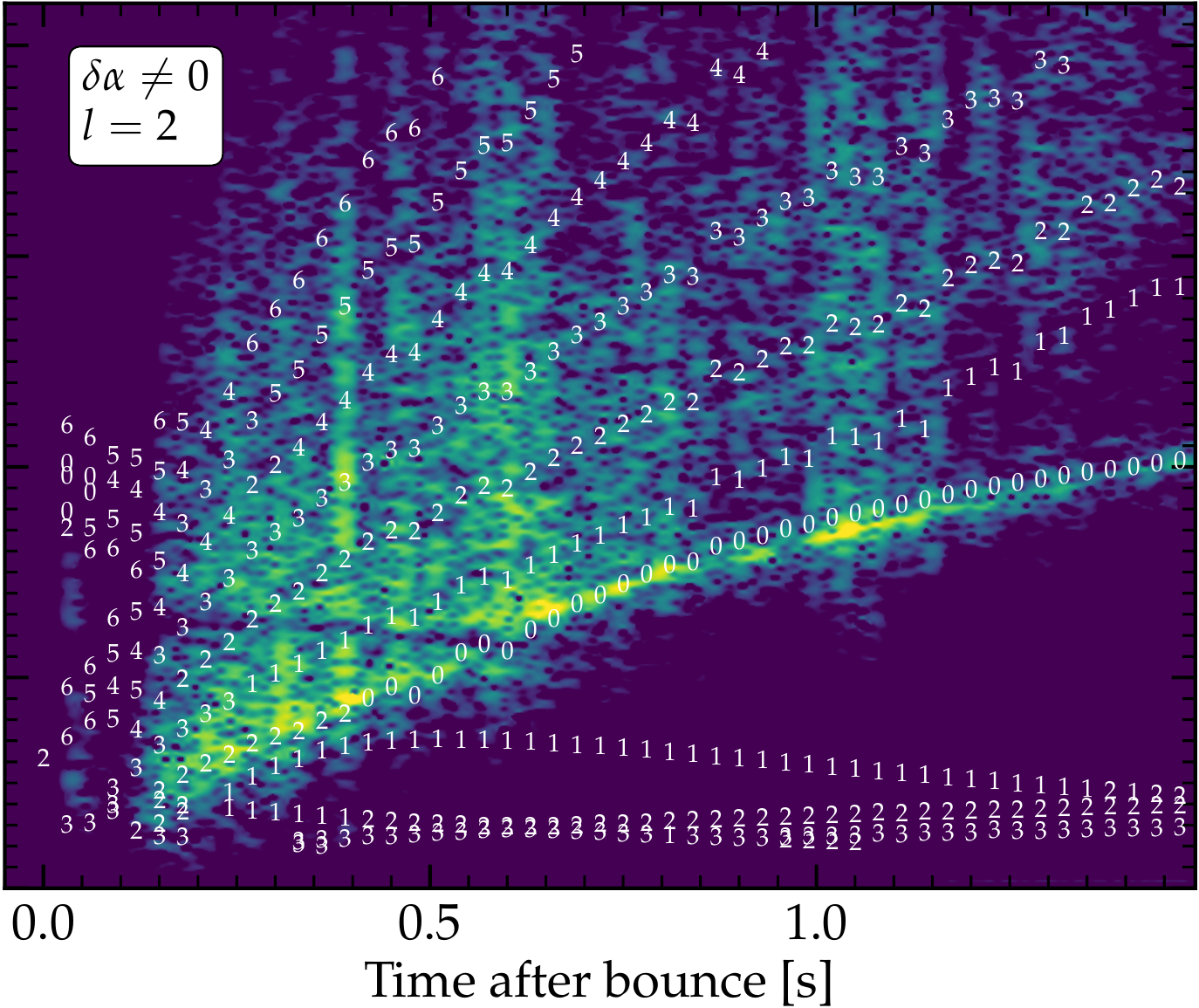}
\caption{Spectrogram for a 10 M$_\odot$ non-rotating progenitor derived from an axisymmetric simulation using the FORNAX code and the SFHo nuclear equation of state, taken from  \citet{MoRaBu18}. Superimposed on the spectrogram are the results of a 
modal analysis for $l=2$ modes. The peak frequency emission is tracked early on by a $g^{2}_{2}$ mode and later on by an $f_{2}$ mode. The results were obtained by relaxing the Cowling approximation (i.e., $\delta\alpha\ne0$).}
\label{fig:gfmode}
\end{figure}

The results from two-dimensional CCSN models pointed to surface $g$-modes excited by accretion funnels from the gain layer impinging on the PNS surface. Three-dimensional models on the other hand have painted a more complex and varied picture, as discussed in Section \ref{sec:sourcing}. $g$-modes can be excited by accretion onto the PNS and by convection (specifically, convective overshoot) deep within the PNS. In both cases, the relationship to the peak frequency, Eqn. (\ref{eq:gmodepeak}), remains the same, but $M$ and $R$ are obviously not the PNS mass and radius in both cases. With regard to the fundamental mode, this ambiguity does not arise given that $M$ and $R$ in Eqn. (\ref{eq:fmodepeak}) are in fact the PNS mass and radius. However, up until now we have assumed that the peak frequency emission seen in CCSN spectrograms is due to a single mode, $g$ or $f$, but given the results of past analyses, this emission results from a combination of $g$- and $f$-modes, with $g$-mode emission dominating early and $f$-mode emission dominating late. (See Figure \ref{fig:gfmode}, which is taken from \citet{MoRaBu18}.) 

{\em Thus, our ability to cull information about the PNS from a CCSN GW detection is predicated on (i) capturing the evolution of the peak frequency of the GW emission, (ii) determining which modes are present in the peak emission, and (iii) sourcing the peak emission.} 

Of course, the surface of the PNS provides the natural boundary for any modal analysis, but it is important to note here that different groups define this surface to be at different densities -- e.g., $\rho=10^{10}$ or $10^{11}$ gcm$^{-3}$. To further complicate matters, \citet{SoKuTa19} have demonstrated that the outcomes of the modal analyses are sensitive to the choice of outer boundary. And regarding (ii), \citet{RoRaCh23} have recently shown that the classification of modes varies as the method of classification varies, as well. That is, in one classification a mode may be classified as a $g$-mode, whereas in another, it may be classified as an $f$-mode. 

Other difficulties include the fact that modal analyses of PNS GW emission have been confined to spherical symmetry (i.e., the analyses are conducted on spherical averages of three-dimensional simulation data). Aside from the obvious implications of averaging the simulation data, the inclusion of rotation in the analysis is not possible without breaking the assumed symmetry, which would in turn render the modal analyses far more challenging. As demonstrated and discussed by \citet{PoMu20}, one might then expect that analytic expressions for the evolution of the peak frequency derived under the assumption of spherical symmetry may no longer accurately track the predicted evolution of the peak frequency from CCSN simulations, particularly as the progenitor rotation increases.

Finally, a more accurate determination of the modes that may exist in the PNS and their frequencies is made possible when the equations that are linearized are the equations that are used in the CCSN simulations providing the modal analyses backgrounds. A number of the models that have been used for such modal analyses, including the models discussed here, deploy the pseudo-Newtonian treatment of gravity discussed in Section \ref{sec:Model Status}. On the other hand, the majority of the modal analysis tools that have been developed for the purpose of PNS asteroseismology are based on the conformally flat condition (CFC) implementation of general relativity. This may lead to over-estimation of the mode frequencies, as discussed extensively by \citet{Westernacher-Schneider20}. Moreover, the results shown in Figure \ref{fig:gfmode} were obtained by relaxing the so-called Cowling approximation, wherein the metric variables (the lapse function, $\alpha$, the shift vector, $\beta$, and the conformal factor, $\psi$) are not varied in the linear analysis. Results using the Cowling approximation are included in \citet{MoRaBu18}, as well, and yield different results not in sync with the peak frequency evolution, for any mode. Further discussion on the validity of the Cowling approximation can be found in \citet{ToCePa19} and \citet{SoTa20}.

\vspace{-12pt}
\section{Detection and Parameter Estimation of Gravitational Waves from Core Collapse Supernovae}

Up to this point we have reviewed (a) the status of the numerical modeling of CCSNe and the challenges of achieving accurate modeling of the explosions and (b) the relationship between the physics of the central engine and the GW production. The extraction of physical information from GWs requires that consider we investigate (c) the response of a laser interferometer to the GWs from CCSNe, as well as multimessenger opportunities/challenges. This sector of CCSN research could be divided into (a) detection strategies, (b) PE strategies, and (c) the role of the temporal observation window on (a) and (b).

\vspace{-10pt}
\subsection{Detection Strategies}

In this section we describe the key conceptual elements involved in the detection process 
for GWs from CCSNe and the subsequent PE. In order to understand the performance and mathematical structure of the current detection and PE algorithms, it is useful to review the time-frequency properties of the data recorded by laser interferometers and their relationship with related algorithms for CBCs. 

Laser interferometers are sensitive to a linear combination of the two GW polarizations with additive noise
\cite{Misner:1973prb}:
\begin{equation}
    x_i(t)= h_i(t)+n_i(t),
\label{eq:recordedx}
\end{equation}
with
\begin{eqnarray}
    h_i(t) & = & H_{i+}(\theta,\phi, \psi) h_+(t-\tau_i) \nonumber \\ 
           & + & H_{i\times}(\theta,\phi, \psi) h_\times(t-\tau_i),
\label{eq:ifo_response}
\end{eqnarray}
where $\tau_i$ is the delay of the interferometer $i$ with respect to a fixed, Earth location, usually the arrival time at the center of the Earth, and $\theta$, $\phi$ and $\psi$ are Euler angles in the interferometer reference frame having the two arms located as its $x$ and $y$ axis. See \citet{Vitale:2011zx} for a review of the relationships between the three types of reference frames involved in the interferometer responses: the interferometer frames, the Earth frame, and the GW frames. $H_+$ and $H_\times$
are the interferometer responses (or antenna patterns)
\begin{eqnarray}
  H_{i+}(\theta,\phi, \psi) & = & \frac{1}{2}(1+\cos^2(\theta))\cos 2\phi \cos 2 \psi \nonumber \\
                            & + & \cos \theta \sin 2 \phi \sin 2\psi , 
\label{eq:ifo_response_H+}
\end{eqnarray}

\begin{eqnarray}
    H_{i\times}(\theta,\phi, \psi) & = & -\frac{1}{2}(1+\cos^2(\theta))\cos 2\phi \cos 2 \psi \nonumber \\
    & + & \cos \theta \sin 2 \phi \cos 2\psi , 
\label{eq:ifo_response_Hcross}
\end{eqnarray}
and $\psi$ is a gauge-freedom angle determining in the Cartesian frame whose $z$ axis is the direction of propagation of the GW how to fix the GW $x$ and $y$ axis and consequently $h_+(t)$ and $h_\times(t)$. $h_i(t)$, which is gauge invariant, varies, however, depending on the interferometer location, orientation, and hardware specifics. It is worth specifying that the exact interferometer response is frequency dependent and the generally-used equation above is a long wavelength approximation carrying errors of at most 1 to 2 percent for frequencies up to a thousand Hz \cite{Rakhmanov:2008is}. 
%To provide intuition on how the response is frequency dependent, at GW frequencies where the period is equal to the time a photon stays in the cavity, the response of the instrument is zero.

\begin{figure*}
\includegraphics[width=0.49\textwidth]{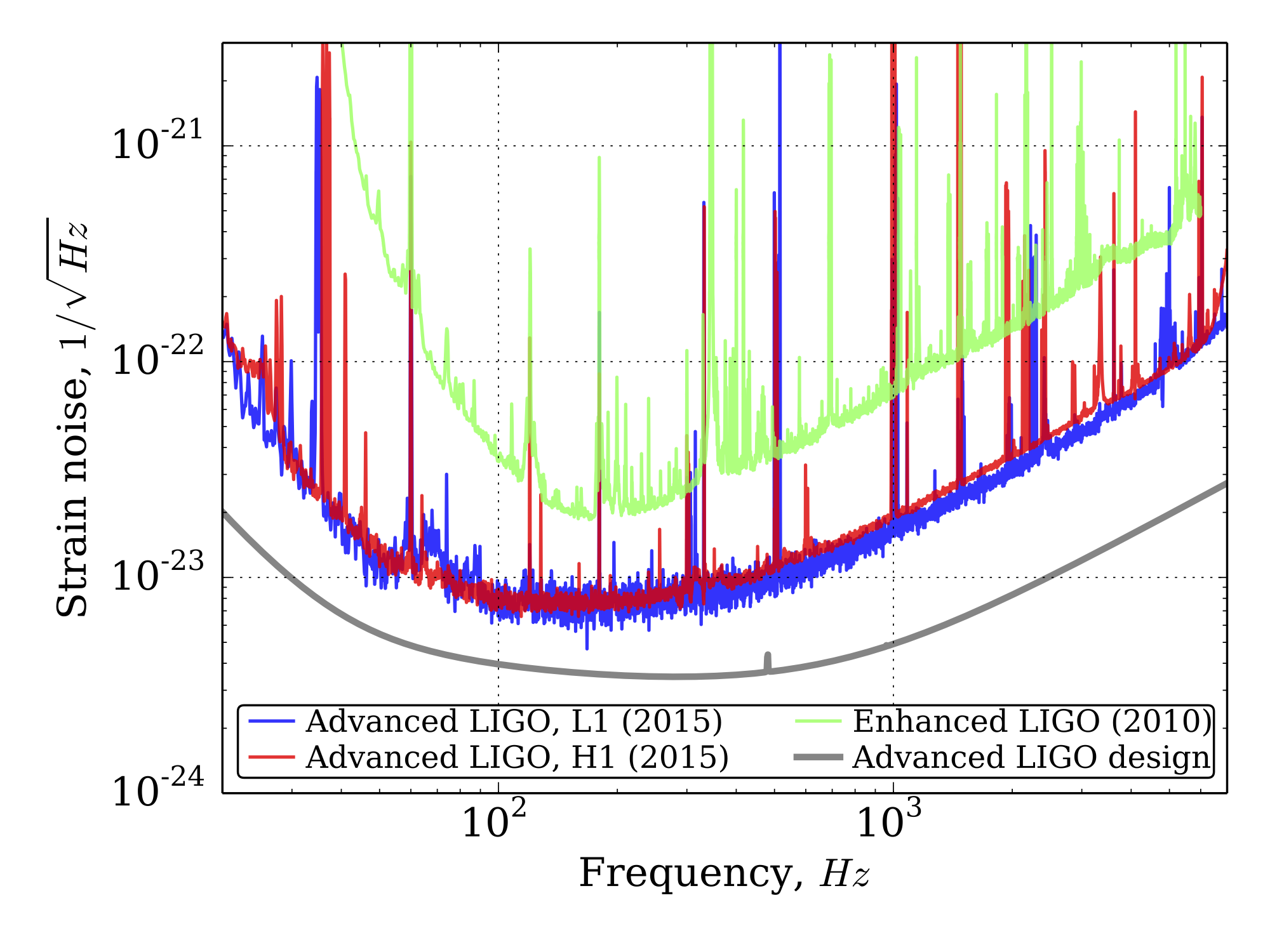}
\includegraphics[width=0.49\textwidth]{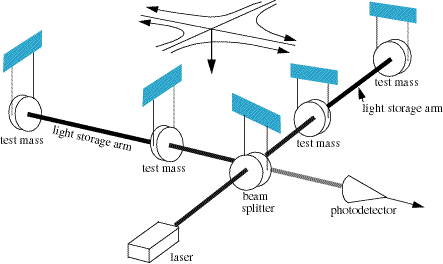}
\caption{Interferometer diagram and examples of noise spectra. The four main features defining the noise spectra are (a) sharp increase of the noise at low-frequency due to ground vibrations (b) the increase of the noise at high frequencies because of quantum effects like photon shot noise (controlled by Posson statistics) (c) the lower limits imposed by the coating vibrations in the most sensitive frequency and (d) the spikes related to resonant frequencies of the apparatus (especially mirror supporting strings), calibration frequencies "lines", and power lines disturbances (60Hz in the US). The interferometer geometry has a natural choice of reference frame where the two arms are labelled x and y arms. }
\label{fig:ifostrain}
\end{figure*}

The detection process involves computing detection metrics from the data and declaring a detection if the values are above particular thresholds chosen to limit the False Alarm Probability (FAP; usually for GW CBC signals a 5-sigma FAP is used). We want algorithms that maximize the detection probability for a desired maximum tolerable FAP. In other words, we want the best possible Receiver Operating Curve (ROC). It is also useful to mention the concept of statistical bias in designing an algorithm. While there is a large amount of freedom in defining the strategy, one is not allowed to choose the detection architecture and related thresholds after looking at the GW candidate events (commonly described as ``opening the box"). The reason is, one could make decisions that minimize the FAP for an event of interest, which would not have been made just looking at noise events (the so-called time-lag procedure) or GWs from  injected CCSN (see for example \citet{Szczepanczyk:2023ihe} and references therein). 

In the following we review the issues involved in choosing the best algorithms for the detection and PE of CCSNe. The guiding principle is to incorporate as much as possible the idiosyncrasies of the instrument response to GWs, noise, and GW morphological features, to maximize the ROCs and PE accuracy.

According to the Neyman-Pearson lemma,  the metric producing the best ROC for a scalar detection problem where a single measurement x is performed in the presence of additive Gaussian noise is the so-called likelihood ratio \cite{Neyman_pearson:1933}. This is the ratio of the Probability Distribution Function (PDF) of the measurement if the signal is present to the PDF if it is not:
\begin{equation}
    \Lambda=\frac{p(x(t); h(t,
    \xi))}{p(x(t))}\,,
\label{eq:likratio}
\end{equation}
where $h(\xi)$ represents the dependence of the measurement on a function of astrophysical parameters $\xi$. Such ratios can also be calculated for measurements of time series in Gaussian noise and a network of laser interferometers. Unfortunately there is no theorem indicating the optimal metric for a realistic network of laser interferometers (and in general, non-Gaussian noise). With the assumption that even in the absence of a general theorem supporting the usage of the likelihood ratio, which is generally given by 
\begin{equation}
    \Lambda=\frac{\Pi_{i=1}^{K}p_i(x_i(t); h_i(t,
    \xi))}{\Pi_{i=1}^{K}p_i(x_i(t))} ,
\label{eq:likratio_network}
\end{equation}
can still be used as a starting point in designing a detection algorithm, although with some customizations, which we review in this section. The index $i$ runs through the different interferometers of the network, $p_i$ are the PDFs of the data at different interferometers, which, for non co-located interferometers, can be assumed to be statistically independent (this is the reason why the numerator and denominator are written as products of the probability distributions at different sites).  

The PDF of each time series at the interferometer $i$, when non-Gaussian noise disturbances (glitches) are not present, can be written as a time-invariant, correlated-Gaussian process
\begin{eqnarray}
 & & p(x,h) = \\ \nonumber
        &   & Me^{-\int_{-\infty}^{+\infty}(x(t_1)-h(t_1))R(t_1-t_2)(x(t_2)-h(t_2))dt_1dt_2},
\label{eq:one_t}
\end{eqnarray}
where $R(t_1-t_2)=\langle n(t_1)n(t_2)\rangle$ is the noise correlation function (here we denote by $\langle . \rangle$ the process of statistical averaging) and the normalization M is signal independent. If we discretize the data and perform a discrete Fourier transform of the data (to illustrate the concepts we assume that we collect N temporal samples at each inteferometer):
\begin{equation}
\tilde{x}(m)=\sum_{l=0}^{N-1}e^{-\frac{2 \pi lm}{N}}x(t_l)\equiv\sum_{l=0}^{N-1}F_{ml}x(t_l),
\label{eq:twoA}
\end{equation}
Eqn. (\ref{eq:one_t}) becomes 
\begin{equation}
p(\tilde{x},h)=\tilde{M}e^{-(\tilde{x}-\tilde{h})^\dagger C(\tilde{x}-\tilde{h})},
\label{eq:twoB}
\end{equation}
where the covariance matrix, whose elements are given by $C_{lp}=(FRF^{-1})_{lp}$ in the frequency domain, is diagonal, with eigenvalues equal to 1 over the square of the noise spectra at the specific frequency (see the curves in the left panel of Figure \ref{fig:ifostrain}, and Eqn. (2.4) of \citet{Sutton:2009gi} for a detailed proof). If the data are whitened (see the discussion later in this section), $C_{ll}$ become identical for all values of $l$. Applying thresholds on $\Lambda$ or $L=\log (\Lambda)$ is equivalent, with the benefit that $L$ can be written as 
\begin{equation}
 L=\frac{1}{2} \sum_{i=1}^{K}  -(\tilde{x_i}-\tilde{h_i})^\dagger C^i(\tilde{x_i}-\tilde{h_i}),
\label{eq:three}
\end{equation}
where we sum over the $K$ interferometers of the network ($C^i$ is the covariance matrix of  interferometer $i$), and we omit the additive signal-independent terms in $\Lambda$, like the logarithm of the normalizations, given they are constants. Calculating $\Lambda$ from observed data runs into the problem that $h_i(t)$ are unknown. In the case of CCSNe we do not have analytical templates describing the expected GWs. The calculation of $h_+(t)$ and $h_\times (t)$ can be performed by imposing stationarity of $L$ (similar to the derivation of Lagrange's equations from a classical action) either in the time domain or in the frequency domain: 
\begin{equation}
 0=\frac{\delta L}{\delta h}|_{h=h_{\rm stationary}}.
\label{eq:four}
\end{equation}
For CBC signals, solving Eqn. (\ref{eq:four}) and plugging the solution into $\Lambda$
is equivalent to computing the so-called matched-filter signal-to-noise ratio (SNR) \cite{Creighton:2011zz} at each interferometer, 
\begin{equation}
    {\rm SNR}_{\rm MF}=\sqrt{\int_{-\infty}^{+\infty}\frac{h(f)^2}{n(f)^2}df} ,
\label{eq:snr_mf}
\end{equation}
where $h(f)$ is the strain spectrum recorded at the interferometer and $n(f)$ is the noise spectrum at the same interferometer. This SNR is an important metric for detectability because it measures the ratio of the squared spectrum of the signal with respect to the squared spectrum  of the noise, at each frequency (this expression suggests that the low-frequency seismic noise in ground interferometers will pose a challenge in achieving large SNR values with the low-frequency GW memory from CCSNe). For template searches, calculating the ${\rm SNR}_{\rm MF}$ involves performing cross correlations of waveform templates with the interferometric data and choosing the template that (on average among different interferometers) best correlates with the data (assuming that the data contains a GW). The waveform templates depend on three types of parameters:  physical (such as the mass of the two compact objects), geometrical (such as the direction of arrival), and those corresponding to testing GR (such as effects due to possible new polarizations). The calculation of the cross correlation for CBC signals is performed by varying all the parameters at the same time and choosing the values that maximize the correlation both for detection and PE. The dimension of the parameter space is constrained to up to a few tens of parameters. 

For burst signals, and CCSNe in particular, we treat each frequency (or time) sample of $h_{+}(t)$ and $h_{\times}(t)$ as free parameters, and, if there are no constraints, there are  approximately 16384 (number of samples per second), times 2, degrees of freedom for each second of signal. The direction of arrival can be unknown as well (for example in the case where we have a failed CCSN without an EM counterpart and without a neutrino signature as well). 

According to Eqn. (\ref{eq:ifo_response}), $h_{+,\times}(t)$ have identical temporal shape for all of the interferometers, but a specific time delay at each interferometer (see for example the expressions in \citet{Markowitz:2008zj}) for a fixed direction of arrival with respect to a fixed frame of reference whose origin is at the center of the Earth. In general, these delays are unknown. Recalling that a shift in time in the time domain corresponds to multiplication by an exponential in the frequency domain: If we know the direction of arrival we can evaluate the $\tau_i$ of Eqn. (\ref{eq:ifo_response}), remove the effect of the delay by shifting the data accordingly in the time domain, or by multiplying by $e^{-2 \pi \tau_i f_l}$ in the frequency domain, and then carry on with the rest of the procedure described below. If we do not know the direction, we can repeat the process below for a grid of sky directions and use the values of the detection metric to estimate the direction. We can also encounter the scenario where we know the direction but the network has poor directional resolution. In this instance, for detection purposes we can employ a grid of sky locations, as well.

For a given direction of arrival, the solution can be obtained at one frequency at a time, of index $l$, to obtain $(\tilde{h}_{\rm stationary}(l))^T=((\tilde{h}_{+})_{\rm stationary}(l), (\tilde{h}_{\times})_{\rm stationary}(l))$. Then, an inverse Fourier transformation to the time domain can be performed to obtain $((h_{+})_{\rm stationary}(t), (h_{\times})_{\rm stationary}(t))$. The solution is usually presented for the scenario where the recorded additive noise has been whitened (see a discussion of the whitening procedure later in this section). Whitening is equivalent to dividing $(\tilde{x}(l))^T=(\tilde{x}_1(l), \ldots , \tilde{x}_K(l))$ by the standard deviation of the noise, $\sqrt{C_{ll}}$, at the corresponding frequency and interferometer, which yields 
\begin{equation}
(\tilde{d}(l))^T=(\frac{\tilde{x}_1(l)}{\sqrt{C_{ll}^{1}}}, \ldots , \frac{\tilde{x}_K(l)}{\sqrt{C_{ll}^{K}}}). 
\label{eq:d}
\end{equation}
The corresponding solutions of Eqn. (\ref{eq:three}) at a fixed direction of arrival (see for example \citet{Sutton:2009gi} and \citet{Klimenko:2015ypf}) are
\begin{equation}
\tilde{h}_{\rm stationary}=(\tilde{H} \tilde{H}^T)^{-1}\tilde{H}\tilde{d},
\label{eq:stationary}
\end{equation}
where
\begin{equation}
\tilde{H}_{+}=(\frac{H_{1+}}{\sqrt{C^{1}_{ll}}}, .., \frac{H_{K+}}{\sqrt{C^{K}_{ll}}}),
\label{eq:Hplustilde}
\end{equation}
\begin{equation}
\tilde{H}_{\times}=(\frac{H_{1\times}}{\sqrt{C^{1}_{ll}}}, .., \frac{H_{K\times}}{\sqrt{C^{K}_{ll}}}),
\label{eq:Htimestilde}
\end{equation}
and
\begin{equation}
{\tilde{H}}=
\begin{bmatrix}
{\tilde{H}_{+}}  \\ \tilde{H}_{\times}
\end{bmatrix},
\label{eq:Htilde}
\end{equation}
and where, in the antenna patterns, the $\theta, \phi$ angles have to be individually calculated, for the chosen direction of arrival, with respect to a fixed, Earth reference frame.   

Inspection of the inversion in Eqn. (\ref{eq:stationary}), meaning that  
$(\tilde{H} \tilde{H}^T)^{-1}$ is real and symmetric and that if we increment $\psi$ by an amount $\gamma$ the following relations hold:
\begin{equation}
\tilde{H}_{+}^{'}=\cos\gamma \tilde{H}_{+} +\sin\gamma\tilde{H}_{\times}
\label{eq:rot1}
\end{equation}
and
\begin{equation}
\tilde{H}_{\times}^{'}=-\sin\gamma \tilde{H}_{+} +\cos\gamma\tilde{H}_{\times},
\label{eq:rot2}
\end{equation}
reveals that at each frequency there is a choice of $\psi$, and consequently $h_{+,x}(t)$, where $(\tilde{H} \tilde{H}^T)^{-1}$ is diagonal, the network has the maximum antenna response  to one of the two polarizations, and, in an orthogonal direction, the minimum antenna response. 
In the reference frame corresponding to the value of $\psi$ where $(\tilde{H} \tilde{H}^T)^{-1}$ is diagonal, the inversion is performed by two operators $\frac{(H^{'}_{+})_{l}}{|H^{'}_{+}|_{l}}$ and  $\frac{(H^{'}_{\times})_{l}}{|H^{'}_{\times}|_{l}}$ that project and rescale the data of the interferometers. 
The full inversion is not possible in the case of aligned interferometers that are blind to one of the two polarizations. See Eqn. (2.31) of \citet{Sutton:2009gi} and the corresponding discussion in \citet{Klimenko:2015ypf} for the use of the ratio 
$\frac{|H^{'}_{+}|}{|H^{'}_{\times}|}$ as a tool to assess the relative sensitivity of a network to the two polarizations in the primed frame and the discussion on constraints (sometimes termed regulators) later in this section for how this information can be used in the design of the detection/reconstruction algorithms. This frame is named the Dominant Polarization Frame (DPF) \cite{Klimenko_2005,Klimenko:2015ypf,Sutton:2009gi}. It is important to stress that the exact value of $\psi$ that diagonalizes $(\tilde{H} \tilde{H}^T)^{-1}$ depends on the orientation of the interferometers, as well as on the relative amplitude of the noise spectra between different interferometers (if the amplitude of the spectra at all interferometers is the same, the quantitites $\sqrt{C^{i}_{ll}}$ factor out in Eqns. (\ref{eq:d}) to (\ref{eq:Htilde}) making the $\psi$ of the DPF frequency independent). When this relative amplitude varies with frequency, so does $\psi$. 

Interferometric data is available for periods of time that can reach days at a time. However, the data is divided into shorter intervals (candidate events) of up to a few seconds, processed with wavelet decomposition and clustering, described later in this section. The resulting reconstructed signals from applying the procedure above to the candidate events are then plugged into L giving:
\begin{equation}
 L_{\rm stationary}=\sum_{l} \left[ \frac{(H^{'}_{+}\tilde{d})_{l}^{2}}{|H^{'}_{+}|_{l}^{2}}+
 \frac{(H^{'}_{\times} \tilde{d})_{l}^{2}}{|H^{'}_{\times}|_{l}^2}\right] =
 \sum_{l} {\rm SNR}^2_l.
\label{eq:totalsnr}
\end{equation}
This quantity can be calculated for an arbitrary direction of arrival, and only the largest value of the values obtained from all directions is retained as the representative value for an event. Histograms of the $L_{\rm stationary}$ values for all events can be produced for the data that are processed in a specific CCSN search or observation temporal window (see \ref{eq:MMA_section}). If only the most energetic event is considered for detection, we refer to the method as the Loudest Event statistical method. 

If the Loudest Event method produces a FAP that is well above what is needed to declare a detection, other approaches must be considered. In the course of a GW search in which we will assume for the moment that the detection of GWs from a CCSN has occurred, there will be many noise-induced events as well. Moreover, with regard to the physical signals detected, there could be more than one -- e.g., in the case of black-hole formation in a failed supernova. The CCSN-associated signal (successful or failed) could be the loudest event, the weakest event, or neither. In the case of multiple candidate events, ``population studies methods'' can be triggered, in which one looks for evidence of the CCSN given the distribution of the number of events as a function of SNR. These techniques, already employed for the search of GWs from gamma-ray bursts (see for example \citet{LIGOScientific:2021iyk}) focus on detecting the presence of GWs in a population of events that can also contain noise events. Population studies have the potential to establish detection using events that would not be detectable with the Loudest Event statistics approach. It is worth noting, this approach is similar to the approach used for the detection of the Higgs Boson. In the energy distribution of the recorded events at CERN, the Higgs was detected not because it produced the most energetic events, but because the observed distribution profile versus energy would not be expected in the absence of the Higgs.

In either scenario it is necessary to produce populations of noise-only events, by either processing data outside the designed observation window (off-source data) or by changing the time stamp of the data in one or more of the interferometers by an amount much larger than the distance between interferometer sites divided by the speed of light (producing so-called time-lagged data). These noise populations allow us to establish the FAP of the candidate events (see the discussion in section \ref{eq:MMA_section}).

A recent application of this procedure for CCSNe can be found in \citet{Szczepanczyk:2023ihe} (this reference also summarizes the current detection ranges for CCSNe, as well as provides some conservative projections for the near future). However, the orders of magnitude more degrees of freedom relative to CBC signals and the intrinsic weaker GW emissions affect, in comparison to CBC signals, the detection range, as well as the ROCs. Years of trials and code development demonstrated that the detection performance is much better if the projectors are not applied directly to the interferometric data but, instead, to pre-processed data, with modules that isolate the more relevant degrees of freedom for CCSNe and, in general, burst searches. The key ingredients we describe below in optimizing the usage of the likelihood ratio for CCSNe are the pre-processing modules (1) whitening, to remove predictable components of the noise, (2) wavelet decomposition and clustering, to consider only connected time-frequency structures of realistic duration for GWs from CCSNe, (3) network constraints, to include the relative sensitivity to the two GW polarizations and the relative sensitivity in different sky locations (this and the next category have the goal of reducing the rate of false alarm events), and (4) morphological constraints, such as duration constraints or time-frequency contraints. Furthermore, in the search and PE we also apply (5) vetoes and data quality, to eliminate events with strong statistical correlation to Earth-induced noise sources, and in this review briefly discuss the difference between (6) frequentist versus Bayesian approaches to incorporating constraints.

The gap in detection performance between CBC and CCSN algorithms and signals with equal ${\rm SNR}_{\rm MF}$ is decreasing with time through these improvements. Several algorithms have been used in this effort over the years (see for example TFClusters \cite{Sylvestre:2003gm}, Q-pipeline \cite{Chatterji:2004}, X-pipelines \cite{Adams:2013pna}, and cWB \cite{Klimenko_2005,PhysRevD.103.103008}), and more are under development. New trends also involve investigating pure ML algorithms for the detection of GWs from CCSNe.

(1) {\bf Whitening:} It can be shown that the noise of the laser interferometers has components that are predictable given previous temporal samples. It has also been shown that it is easier to detect GWs if these components are subtracted before performing the subsequent steps of the analysis. These components can be estimated with linear predictor filters (LPF; for example with the Levinson algorthm described in \citet{Press:1991}), where the coefficients are determined or ``trained'' with past test data:
\begin{equation}
    x_{i,{\rm predicted}}(t_{n})=\sum_{l=1}^{P} a_l x_{i}(t_{n-l}),\\
\label{eq:white_1}
\end{equation}
where the $a_l$ are the P coefficients of the LPF.  The data actually analyzed becomes
\begin{equation}
    x_{i,{\rm analyzed}}(t_{n})=x_{i}(t_{n})-x_{i,{\rm predicted}}(t_{n}).
\label{eq:white_2}
\end{equation}
This subtraction process also has the effect of whitening the resulting data, from which the name originated. There are several implementations of these LPFs in GW data analysis algorithms. Two that deserve special mention for CCSNe are the implementation in the wavelet domain (see the subsequent discussion of wavelets) performed by cWB \cite{Klimenko_2005}, where the wavelet family is chosen to reduce as much as possible the frequency leakage described below, and the implementation by MULASEC \cite{PhysRevD.103.103008}, where the values of the algorithm's free signal-processing parameters (like those of the adopted Wiener filters and convolutional neural network) were specifically optimized by testing the performance on GWs from CCSNe.  In both situations the whitening has been shown to increase the SNR of the candidate events, because the LPF subtracts energy from the noise but not from from the signal (consistent with the fact that the LPF training is performed on noise and not GWs). These LPFs have to be optimized to account for the fact that the noise spectrum is time dependent and has a drastic frequency variability, especially below 20 Hz. This last fact makes it prone to energy leakage between frequencies. For example, when we analyze a finite-duration signal extracted from a time series, it is equivalent to multiplying an infinite time series with a time window of finite duration. In the frequency domain this is equivalent to a convolution between the spectrum of the original signal and the spectrum of the time window. Every time window of finite duration has a finite nonzero bandwidth, which automatically broadens the band of each of the frequencies in the original signal. If someone is not careful the broadening of the low-frequency noise amplitude could ``fill'' the ``noise spectral well'' around 100 Hz.

(2) {\bf Wavelet Decomposition:} Instead of applying the inversion equations for $h_{+,\times}(t)$ on just recorded time series or whitened time series, it has been shown to be useful to also reduce the degrees of freedom with methods developed for time-frequency representations of the data. The usage of spectrograms has a long history in signal detection and characterization (see for example \citet{Chan_2020} and references therein). Spectrograms are a special case of wavelet decomposition. The detectability of a GW signal increases when the signal can be described with the least possible number of wavelet components (see for example \citet{Sutton:2013ooa} and references therein). The ultimate performance is achieved when a signal is matched by a single wavelet, rendering the search a template search. If someone would have to take inspiration from CBC GW signals and choose an appropriate wavelet decomposition the wavelets would be temporally long and narrowband at lower frequencies and temporally short/broadband at higher frequencies (because the frequency is slowly evolving at the beginning and rapidly evolving near the merger). Having narrowband wavelet components for low frequencies is also useful for controlling frequency leakage from the so-called ``seismic wall'' at low frequencies.

The general approach used for CCSN signal processing is to use CBC-inspired wavelet decomposition with, however, larger dimensionality than the dimensionality of the time series -- for example by using multiple copies of the same wavelet temporally translated by a fraction of the wavelet duration. To illustrate how this could be relevant for GWs from CCSNe, let us assume that we have in the GW a SASI component made of a single frequency of a duration of 100 milliseconds and that we have a wavelet basis containing wavelet components having exactly the same frequency and duration, but in performing the wavelet decomposition, the wavelets are obtained by translating the mother wavelet exactly 100 milliseconds. It is quite likely that none of the wavelets have start and end times exactly identical with the SASI component. The choice of temporally translated wavelets by a {\em fraction} of the duration can aid the detection because the probability of perfect alignment increases and a wavelet perfectly alligned with the GW could produce a wavelet coefficient amplitude that passes internal algorithm thresholds, whereas partially-correlated wavelets might not. See for example the discussions on ``black-pixel'' thresholds in \citet{Klimenko_2005}, where only a small percentage of more energetic wavelet components is retained, as well as only those that form clusters. This last strategy is motivated by the canonical spectrograms from CCSNe -- e.g., see Figure \ref{fig:Mezzacappaetal2023Heatmaps}. The fraction of data retained and the clustering procedure used are some of the tunable parameters of the algorithm. The tuning process is usually based on maximizing ROCs and detection ranges for the average of trials with a representative population of GW signals.

It was shown that using multiple wavelet decompositions in parallel is beneficial \cite{Chatterji:2004}, and now it is common to have this feature in detection algorithms (see for example \citet{Adams:2013pna} and \citet{Klimenko_2005}). We foresee more research being conducted on this front, as the progress has been driven by testing the assumptions on evolving data and CCSN GW databases. 

(3) {\bf Network Constraints:} These involve (a) metrics that quantify the similarity of the reconstructed signal between different interferometers and (b) constraints on polarization (if a configuration of the laser interferometer network is mostly sensitive to a single polarization, we do not try to reconstruct the other because it would produce mostly noise-induced events -- for example by choosing in the middle expression of Eqn. (\ref{eq:totalsnr}) only the contribution from the $+$ polarization). (c) A network may also have sky locations where it is particularly insensitive, and events reconstructed in these directions might also be mostly noise-induced events.

(4) {\bf Constraints on signal time-frequency structure:} In CCSN searches, we assume that the frequency content is limited between $\sim$20 Hz and $\sim$2000 Hz because the detectors are much less sensitive outside this frequency band. We also do not include events that are longer than a few seconds. In recent years it has proven useful to use different versions of ML algorithms that distinguish noise-induced events from events of astrophysical origin \cite{Cavaglia:2020qzz,Morales2020,Antelis:2021qaz,Lopez:2021ikt}.

(5) {\bf Veto and Data Quality:} For all algorithms devoted to detection and PE, it is critical to identify stretches of data that have been damaged by the coupling with non-astrophysical disturbances. Currently there is no CCSN-specialized approach to this. It is worth mentioning that different stretches of data are sometimes used for different purposes. For example if we have an exceptionally interesting GW candidate or an exceptional astronomical event such as a Galactic CCSN we would use the data even if the time associated with the candidate or event is in a period of time when many known correlations exist with Earth-based sources of strong glitches. However, where no event is deemed a GW event, astrophysical statements can still be made -- for example on the maximum energy emitted in GWs for the source of interest (the most recent of such statements for CCSNe were produced in \citet{Szczepanczyk:2023ihe}). These statements are based on the fact that no events louder than the loudest observed noise event were present in the data set. For this usage, it is usually better to discard intervals of the data with transients correlated to known noise disturbances so that astrophysical statements are stronger.

The noise can also be divided into an ever-present Gaussian component (which emerges as a product of many small disturbances blending, as described by the central limit theorem) and glitches -- i.e., sporadic, sub-second transients induced by environmental disturbances (a falling tree, a train, a fluctuation in the power supply, etc.). These glitches, which affect more strongly the detection confidence than PE algorithms, are studied and classified on a continuous basis by different laser interferometer collaborations (see for example \citet{Davis:2021} for LIGO, \citet{Yuzurihara:2023jjv} for Kagra, and \citet{Virgo:2022kwz} for Virgo).

(6) {\bf Frequentist versus Bayesian Approaches:} The use of a frequentist versus a Bayesian approach is one aspect of the algorithms that has both practical and conceptual ramifications. In the frequentist approach the probability depends on the physical value of the parameters treated as deterministic quantities, while in the Bayesian approach the parameters are random variables as well.  In this second scenario the joint PDF can be factorized according to Bayes' theorem
\begin{equation}
p(x,h)=p(x;h)p(h),
\label{eq:bayes}
\end{equation}
where $p(x;h)$ is the frequentist PDF and $p(h)$ is the prior, describing the 
data-independent assumptions someone would like to make regarding the GW, including the direction of arrival. The Bayesian prior needs to be provided. A Bayesian detection process could be implemented in Eqn. (\ref{eq:likratio_network}) by replacing the frequentist PDFs with Eqn. (\ref{eq:bayes}). It may be problematic to make prior assumptions about GW properties before we have detections from CCSNe. In practice though, it should be possible to implement with Bayesian priors the network and morphology constraints discussed earlier in this section. While we do not describe in detail the Bayesian implementation,  \citet{Pannarale:2018cct}, \citet{Sutton:2009gi}, and \citet{Cornish:2020dwh},
and references cited therein, provide a good overview of current efforts.

The strategies in this section have the goal of improving the detection range for a broad range of CCSN morphologies and sky locations. They are constantly re-evaluated because of theoretical and signal processing advancements, as well as the evolution of the instruments, which produce noise spectral properties that are slightly different at every data collection run, including different types of transient disturbances. For a given value of the SNR a matched-filtering approach can still provide a detection range a few times larger than a pure coincident excess-energy approach. However, the techniques in this section are reducing the performance gap. Of course, reducing the number of degrees of freedom of CCSN searches is essential. One important aspect of the entire enterprise we do not cover in this review has to do with the computational cost associated with any given approach adopted. For searches performed specifially for GWs from CCSNe in laser interferometric data, see \citet{Szczepanczyk:2023ihe} and \citet{LIGOScientific:2019ryq,LIGOScientific:2016jvu,LIGOScientific:2009dyd}. 

\vspace{-10pt}
\subsection{Parameter Estimation Strategies}

We now review the main issues involved in PE for CCSNe. A whole range of PE activities have been (or are being) developed for the GW sector and its role in MMA PE (for activities included in CCSN search papers see
\citet{Szczepanczyk:2023ihe} and \citet{LIGOScientific:2019ryq,LIGOScientific:2009dyd} 
and for activities documented in methodological papers see
\citet{LiRiLu23,Gossan:2015xda,Lagos:2023qli,Villegas:2023wsu,Szczepanczyk:2022lnz,Gill:2022amm,Antelis:2021qaz,RiZaAn23,LIGOScientific:2021mwx,Szczepanczyk:2021bkk,Cavaglia:2020qzz,Lin_2020,LIGOScientific:2019hgc,PoMu22,Roma:2019kcd,Suvorova:2019ebd,Pastor-Marcos:2023tcc,DrAnDi23};
more references are discussed in the rest of this section).

The main starting point is the appreciation of the basic features of GWs from CCSNe (with their striking differences relative to other GW transient sources such as CBC sources). Even if the time series display a marked stochasticity, there are deterministic features that are visible either in the time domain, frequency domain, or time--frequency domain. The polarization state is expected to be mostly randomly polarized, with the exception of (a) the core bounce component for rapidly rotating progenitors and (b) spiral SASI effects, which are expected to be strongly and somehow elliptically polarized. We summarize in the following these features and the efforts to both extract them (detect them in interferometric recordings, with algorithms that use minimal assumptions about their morphologies) and quantify them (estimate deterministic parameters, their intrinsic uncertainties, and the expected estimation uncertainties given interferometric data):

(1) {\bf Core Bounce:} The estimation of physical parameters from the GW core-bounce phase of rotating CCSN progenitors aims to both estimate information about the degree of rotation in the progenitor and about the PNS equation of state. It can complement the results of the PE performed on other CCSN GW features (like the evolution of the high-frequency GW emission from the PNS). 

The core-bounce GW component is amenable to be modeled by templates \cite{RiOtAb17}, a special feature, and, accordingly, by techniques already used for modeled searches and their PE, as in the case of matched filtering implemented for GWs from CBCs.

There is active research in choosing the GW templates to match with the GW signal embedded in realistic interferometric noise \cite{Pastor-Marcos:2023tcc,Villegas:2023wsu}. These templates could be sourced directly out of numerical simulations, whose accuracy is evolving over time, or by identifying phenomenological templates. In the second scenario, optimal template construction (using a matched-filtering fitting factor between the templates and numerical simulations as a metric of ``goodness'') can be done in terms of  (a) the size of the parameter space, (b) the types of parameters: phenomenological or physical, such as (i) the ratio of kinetic to gravitational potential energy, $\beta$ (there are indications that the parametrization for  $\beta \leq 0.06$ and above 0.06 are different \cite{RiOtAb17}) or (ii) source orientation, and (c) the GW polarization state. Furthermore, \citet{RiOtAb17} suggest that $\beta$ is more closely related to the core bounce amplitude and $\sqrt{G \bar{\rho}_c}$ to the duration (where $\bar{\rho}_c$ is the average PNS density), with further corrections from different EOS and radial rotational profiles, as suggested for example in \citet{Abdikamalov:2013sta}.

Performing a search and PE in already-detected GW signals from CCSNe
could allow us to detect the presence of the core-bounce component, as well as estimate its physical parameters such as $\beta$. Identification ROCs (meaning assessing the probability to detect the presence of the core bounce for a given false identification probability) can also be produced similarly to what has been developed for the SASI in \citet{LiRiLu23}. The PE of $\beta$ from the core-bounce component could also allow us to establish the presence of rotation if the accuracy is sufficient to establish that $\beta$ is different than zero.

The usage of numerical/phenomenological templates also allows us to study theoretical minima (Cramer-Rao Lower Bounds; CRLB) in the errors any algorithm could achieve for a given signal, interferometer network, and noise floor, as well as higher-order corrections to them when non-linearities in the relationship between the observed data and the estimators of the physical parameters make the CRLB an underestimation of the actual error. The calculation of the CRLB and higher-order asymptotic expansions involve the evaluation of mean values of the log-likelihood derivatives with respect to the parameters. This methodology has been previously applied to GWs from binary black holes (see \citet{Vitale:2011zx} and references therein) to quantify the accuracy in estimating physical parameters such as total mass, reduced mass, and phase. 

(2) {\bf Memory:} The first evidence of GW memory from CCSNe was seen with two-dimensional numerical simulations \cite{MuOtBu09,MuJaMa13,YaMaMe10,YaMeMa15}, even though it was understood that the imposition of axial symmetry would potentially maximize the asphericity of the explosion and, consequently, the memory, given that explosion can occur only along the symmetry axis, in a pronounced prolate fashion. Indeed, in \citet{OCCo18} and \citet{MeMaLa20} the GW strain amplitudes for CCSN simulations initiated from the same progenitor in two- and three-dimensional simulations were compared, with the amplitudes for all phases of the strain in the latter case being much reduced. The key point is, the exact characteristics and asymptotic values of the signal can be determined accurately only from three-dimensional numerical simulations, although it is important to stress that the axial symmetry of rapidly rotating progenitors has the potential to develop much larger memory amplitudes. 

The amplitude of the strain for the low-frequency emission from GW memory can be significantly larger than the strain caused by other sources, such as the high-frequency emission from PNS oscillations or the low-frequency emission from the SASI. However, the energy contained in the low-frequency component is much less because the energy density in the frequency domain is proportional to the square of the frequency.

In the low-frequency limit, GW memory is expected to be linearly polarized. More explicitly, the GW spacetime metric being symmetric and real can always be diagonalized. However the frame where this happens is in general constantly evolving. Because this is not the case for the memory, we can say that it is linearly polarized. The sensitivity of the network to the memory depends on the relative orientation of this frame with respect to the DPF at low frequencies, unless we have a network equally sensitive to both polarizations.

A discussion of the key issues associated with GW memory from CCSNe can be found in
\citet{RiZaAn23}, where they discuss how the GW memory has a peak frequency of approximately 1 Hz, related to the typical time scale for its development, which is approximately one second. The memory dominates the GW emission from CCSNe below a few tens of Hz. The noise floor of ground-based interferometers is currently too high to be able to see the signal above the noise, at 1 Hz. However, the spectrum of the memory decreases with increasing frequency at a lower rate than the noise spectrum. In general, for any ground-based interferometer configuration, a critical feature for memory detectability is the rate of frequency variability of the seismic noise. This rate will evolve as vibration control systems evolve. 

Low-frequency GWs from neutrino emission asymmetries in CCSNe were studied in the context of three-dimensional models by \citet{VaBu20}. They found that the neutrino-generated GW signal from a Galactic supernova will be detectable by AdvancedLIGO, the Einstein Telescope, and proposed space-based detectors. However, this conclusion is for the total signal and not specifically the memory. It is also unclear whether or not the sudden termination of the GW signal at a non-zero value introduced artefacts in the Fourier analysis, or in detectability estimates for present and future ground-based interferometers. The issue of how continuation of truncated simulations should be managed in sensitivity studies performed with real interferometric noise is discussed in \citet{RiZaAn23}, as well as the angular dependence of the matter-induced memory with respect to source orientation. More research is needed for the study of the angular dependence of the neutrino-induced memory amplitude with respect to source orientation. More generally, research is needed to develop a CCSN memory detection pipeline. 

It is worth noting that future space interferometers have strong potential for the detection of GW memory. An initial quantitative analysis is presented in \citet{RiZaAn23}. In addition, \citet{Mukhopadhyay:2021gox} describe how improving the noise of DECIGO by one order of magnitude will allow robust time triggers for CCSNe at distances D $\sim$40--300 Mpc. And \citet{Branchesi:2023sjl} discuss how lunar ranging experiments can be used to detect GW memory, as well. 

(3) {\bf High-Frequency Feature:} This feature is recognizable in a time-frequency spectrogram, as in Figure 3, as a continuous, increasing, approximately linear feature (for the currently visible part of the spectrum, below 1000 Hz), starting at $O(100)$ Hz and increasing up to $\sim1-2$ kHz with time after bounce. The rate of increase of the frequency is believed to be strongly correlated with progenitor characteristics, such as the progenitor mass and degree of rotation (for studies in the context of three-dimensional CCSN modeling, see for example \citet{PoMu20}, \citet{Pan_2021}, and \citet{PaWaCo21}), and with characteristics of the PNS, such as its EOS (again, for studies in the context of three-dimensional CCSN modeling, see for example \citet{KuKoHa17}).

At later stages the frequency evolution is expected to plateau or disappear depending on whether the PNS mass and radius stabilize or a black hole forms. Given current laser interferometers it is expected to be difficult to estimate more than the initial linear growth of this feature because the frequencies above 500 Hz are expected to be mostly buried in the interferometer noise. A first attempt to estimate the slope of the feature with real interferometric noise was performed in \citet{LiRiLu23}, applying a chi-squared method to a low-order polynomial evolution of the resonant frequency. The authors applied the procedure to CCSN events that would be identified by cWB, the flagship algorithm for the detection of GW bursts.  The study in \citet{Bizouard_2021} proposed an approach involving the modal decomposition described in Section \ref{sec:normalmode}, along with a polynomial interpolation and simulated Gaussian noise, to infer the time evolution of a combination of the mass and radius of the compact remnant, as discussed in that section. Currently the best performance in estimating the initial slope with real laser interferometric data is documented in \citet{Lagos:2023qli}, where an optimized neural network approach was used. 

The amount of information that could be extracted from the slope estimates and the corresponding errors (which depend on the signals, the distances, and the interferometer network) in real interferometric data is the focus of current research.

(4) {\bf SASI:} Among the deterministic features in a CCSN GW spectrogram is the feature induced by the SASI. It imparts a distinctive feature to the neutrino luminosities as well, in the form of quasi-periodic fluctuations. This feature makes it a natural candidate for multi-messenger studies. The SASI descriptive parameters include its frequency (in both GW and neutrino channels), duration, amplitude, and GW polarization. The average SASI GW frequency contains information about the average radius of the stalled shock front and the coupling mechanism between the shock wave and the PNS \cite{Blondin:2005wz,Walk:2019miz}. Longer SASI duration could be a distinct feature of failed supernovae \cite{Walk:2019miz}. 

It is of interest to analyze the statistical conditions needed to detect the presence of the SASI and to estimate its parameters for realistic detectors, where noise is present. The noise, as well as the signal processing artifacts, of neutrino and GW detectors are different. However, in both the neutrino and GW channels, noise can induce some energy in the SASI time-frequency regions, thus complicating the analyses.
 
Spectral properties of the SASI-induced features in the neutrino luminosity were described in \citet{Lund:2010kh,Lund:2012vm} and \citet{M_ller_2013} for a specific set of progenitors. The question of the detectability of the SASI was discussed, not with respect to a specific algorithm, but in terms of  the spectral amplitude relative to a Cherenkov detector's shot noise. In \citet{Lund:2010kh,Lund:2012vm}, the shot noise was estimated by Fourier transforming a neutrino time series. The estimated shot noise became independent of frequency. Note, however, that when the statistical fluctuations of the neutrino signals themselves dominate over the noise of the detector's background, the frequency-independence assumption may only serve as a rough approximation. (The neutrino time series used by Lund et al. never dominated the noise.) The extension to a full SASI detection methodology in the neutrino channel was performed in \citet{Lin_2020}, where the ``SASI-meter" was proposed in order to detect the presence  of the SASI with a desired statistical confidence, as well as obtain an estimate of the frequency for the SASI candidates that pass a desired confidence threshold. In that work, it was also pointed out, there is an intrinsic uncertainty in the frequency of the SASI, both in the GW channel and in the neutrino channel, related to the finite duration of the instability.

For the SASI GW emission a few approaches have been explored, each with specific idiosyncrasies. In \citet{Roma:2019kcd} a Bayesian method that uses a training process on an existing database of GWs was proposed in order to identify the presence of the SASI. In this study, magnetorotational emission models were assumed not to contain the SASI, and PE and false identification probabilities were not involved. In \citet{Takeda:2021hmf}, they explored the usefulness of an application of the Hilbert-Huang Transform (HHT). While this is an interesting new direction, the role of noise was not part of the study. In \citet{LiRiLu23}, the work of \citet{Lin_2020} on SASI detectability and PE was extended to real interferometric noise.  The probability that the presence of the SASI can be established was analyzed, as well as the intrinsic uncertainties in determining the SASI frequency and the SASI false identification probability. The method used a frequentist inference, which does not apply prior information of the SASI from any specific numerical simulation. The method used theoretical knowledge to identify conservative boundaries of the time-frequency region of a GW signal where SASI contributions would be present. The scenario adopted is when a CCSN detection has been established in both the (time-coincident) neutrino and GW channels. This work also extends the neutrino analysis in \citet{Lin_2020} with an estimate of the duration of the SASI. The wavelet decomposition of the GW data recorded at laser interferometers \cite{Abbott_2018} was performed using cWB. The main metric that was introduced to detect the presence of the SASI in a GW event was the {\it {relative SASI SNR}}. Explicitly, given a GW event wavelet decomposition, the relative SASI SNR is the sum of the SNR of individual wavelet components in the SASI time-frequency region versus the sum of the SNR of all the wavelet components in the event. This metric is a quantitative measure of the predominance of the SASI time--frequency region with respect to the whole GW event. It is not an overall detection  metric like Eqn. (\ref{eq:totalsnr}).

\vspace{-12pt}
\section{Role of Multimessenger Identification of an Observation Window}
\label{eq:MMA_section}
When searching for GWs from CCSNe it is beneficial for the searches to have information about the time when the GW emission might have reached the Earth. The on-source window (OSW) is a time interval during which we search for the GW transient. It is defined as $[t_1,t_2]$, where $t_1$ and $t_2$ are the beginning and end times, respectively. The main benefit of having a short OSW, $t_2 - t_1$, is that the FAP of a candidate event (the main statistical metric for detection) is given by 
\begin{equation}
    {\rm FAP}=1-e^{-(t_2-t_1) {\rm FAR}}\simeq (t_2-t_1) {\rm FAR},
\label{eq:fap}
\end{equation}
where the first equality is the Poisson probability to have more than zero noise events and the second equality follows from the fact we usually operate when $(t_2-t_1) {\rm FAR} << 1$. An all-sky search and a targeted search are the two types of GW searches we can perform for CCSN GW signals. In an all-sky search the amount of data to be searched corresponds to a duration of the order of one or more years (the whole duration of a scientific data collection run). In a targeted search we can use an EM and/or neutrino observation to search data corresponding to a significantly shorter period of time.

The reduction in the OSW can be used to either (a) lower the FAP at a fixed FAR, (b) increase the FAR at a fixed FAP, or (c) affect a combination of both. The second option makes the search more sensitive because an increase in FAR is achieved by using smaller thresholds in SNR and, therefore, by detecting weaker signals (the SNR of an event is inversely proportional to the distance of the source). The choice of strategy among (a),(b), or (c) depends on whether or not it is more pressing to decrease the FAP, or detect signals at a larger distance, as well as on the properties of the data.

\begin{figure}
\includegraphics[width=\columnwidth,clip]{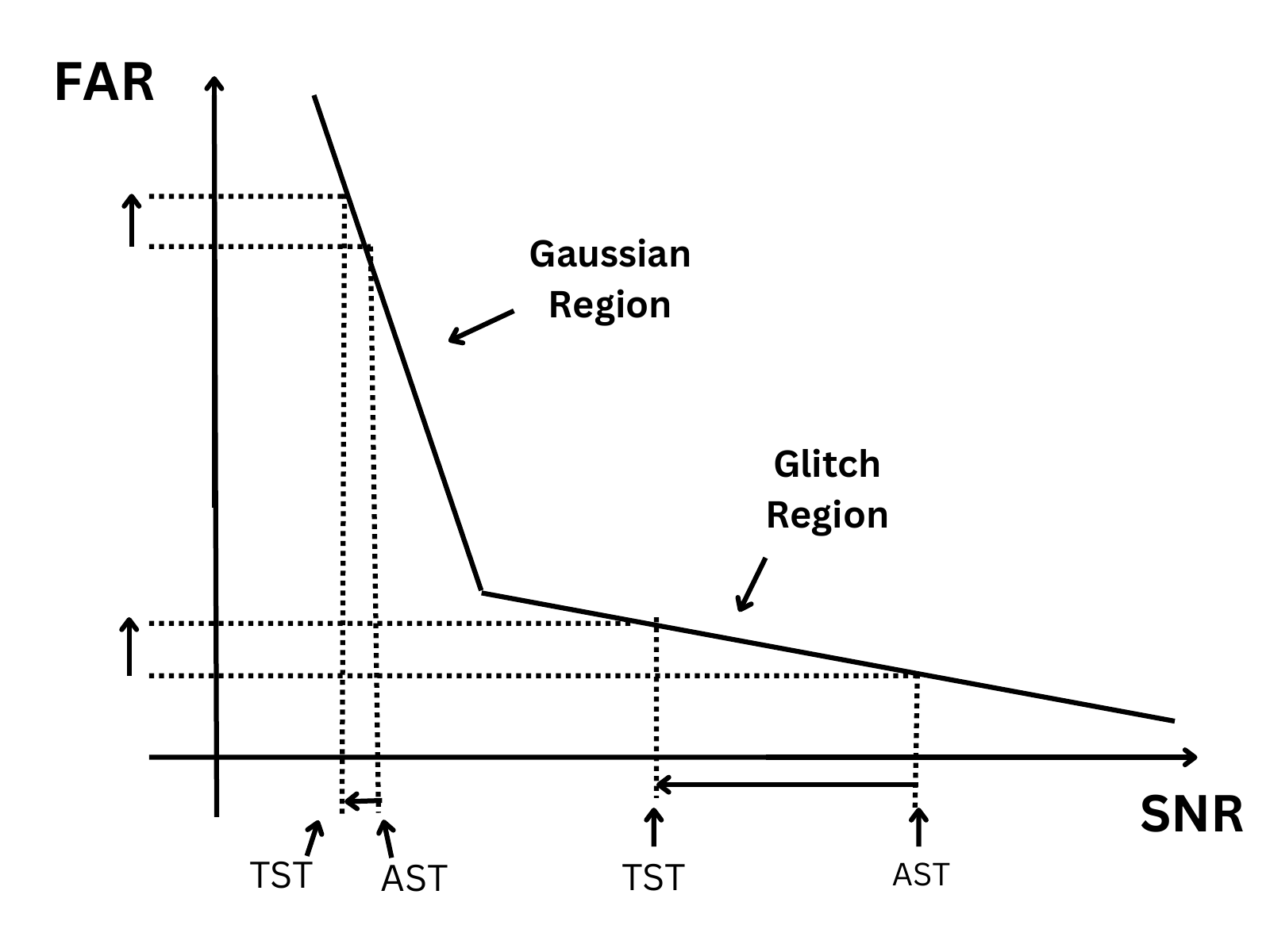}
\caption{The change in the SNR thresholds for a given change in FAR can be different depending on the region of the FAR vs SNR curve, the all-sky threshold (AST), and the targeted-search thresholds (TST). On the vertical axis the two black arrows signify the increase of FAR that we can tolerate while maintaining the FAP fixed because of the reduction of the temporal observation window.}
\label{fig:cartoon}
\end{figure}

The FAR versus SNR distribution, how many noise events per unit time are expected above a given SNR (see for example the cartoon in Figure \ref{fig:cartoon} or Figure 3 of \citet{LIGOScientific:2007hnj} and Figure 5 of \citet{LIGOScientific:2010ped} for real data distributions), might make it easier to get the benefits in (a) rather than the benefits in (b), or vice versa. Histograms of FAR versus SNR can exhibit two types of behavior. At low SNR, the FAR changes very rapidly, by orders of magnitude, for small changes in SNR. At large SNR, when vetoes or other measures fail to remove some of the loud glitches, small changes in the FAR are connected to large variations in the SNR. As a consequence, if the all-sky SNR threshold is in a glitch region, the tolerable increase of the FAR of a targeted search is better used to increase the range, given that the tolerable increase corresponds to a significant decrease in SNR and, consequently, a significant increase in the range. On the other hand, at low SNR, it is better to use the reduced OSW duration to decrease the FAP while keeping the FAR fixed.

The methodology to compute $t_1$ and $t_2$ should also be a compromise between achieving a high probability (see \citet{Gill:2022amm} for some of the limitations in achieving a 100\% probability) to contain the possible GW emission from the CCSN and, at the same time, not lose the benefits of having a shorter OSW relative to all-sky searches (see for example \citet{KAGRA:2021tnv}). 

Before we discuss the OSW options in more detail, it is useful to compare detection and PE scenarios. Detection algorithms in general have multiple internal thresholds that are chosen to identify a GW while discarding with a desired FAP Earth-generated disturbances. Regarding the OSW impact on PE, the situation is different. PE can be performed either on all candidate events (possibly as a way to reduce further the FAP if we use the feedback of PE to disregard events with morphology incompatible with a CCSN) or only on events that have been detected as GW events. While in the first scenario the duration of the OSW will affect how many events will be processed and a characterization of the PE output on the corresponding population of noise events will be needed, in the second scenario the OSW has no direct impact on the PE results. This means, if PE routines are performed in a follow-up mode of already detected events, we can investigate algorithm configurations that would not be viable for detection purposes as long as they legitimately improve the reconstruction performance (see the discussion on SASI reconstruction in the previous section). 

{\bf Neutrino OSW:} The electron-neutrino burst and the rapid rise of the three-flavor neutrino and antineutrino luminosities are expected to happen within $O(10)$ ms of core bounce. The OSW duration can be set to be of the order of a few seconds to accommodate all phases of the anticipated GW emission and possible biases on the GW estimated arrival time, as well as signal start and end times due to performing wavelet decompositions and clustering in the presence of interferometric noise.  In this scenario there are about 7 orders of magnitude improvement that can be achieved in the FAP and/or the FAR (and according to the discussion of this section, sensitivity) with respect to an all-sky search. Studies for extragalactic CCSNe also exist -- e.g., see \citet{IceCube:2014yxu} and \citet{Mukhopadhyay:2021gox}, the latter of whom discuss how a nearly background-free sample of $\sim$3--70 neutrino events per Mt per decade of operation will be available.

{\bf EM OSW:} If neutrinos are not available, measurements of EM emission could be employed as well, as described below. The OSW estimation from EM observations requires two steps: estimating the time of shock breakout (SBO)  -- i.e., the time when the supernova shock wave reaches the progenitor's surface -- and estimating the delay between the GW emission and the time of SBO.

{\em Delay between Collapse and Shock Breakout:} Published work estimating the time delay between the iron core collapse and SBO for a given CCSN progenitor can be found in \citet{Barker:2021iyr}. In particular, their Figure 6 relates the mass of the progenitor to this delay. In addition, an extensive study of CCSN progenitor masses in the local universe was performed in \citet{Smartt:2009zr} (within 30 Mpc, where this study is feasible). As summarized in their Figure 6, in the local universe there is an apparent upper limit of CCSN progenitor masses of 16.5 M$_\odot$. In the absence of progenitor-mass information, the mass--delay relationship in \citet{Barker:2021iyr} would indicate that $\Delta t_{GW-SBO}=3$ days would be an appropriate upper limit for the collapse-to-SBO delay. Clearly then, one discussion that goes beyond the scope of this review but is important for the OSW estimation is the degree of trust we have in the estimate of the CCSN progenitor mass, both for the collapse-to-SBO delay and for the CCSN population characterization. To illustrate the difficulties, in the case of SN2023ixf \citet{Pledger:2023ick} suggested that the progenitor mass was 8 to 10 $M_{\odot}$, \citet{Kilpatrick:2023pse} predicted the mass to be 11 $M_{\odot}$, and \citet{Soraisam:2023ktz} instead predicted the mass to be $20 \pm 4$ $M_{\odot}$.

{\em Estimating the Time of Shock Break Out:} There are three different types of methods that can be used to calculate the time of SBO. The choice depends on the availability of early observational (pre-peak luminosity) data and the availability of published tools to model the light curves.

(i) The Early Observation Method:
In this method, the initial supernova observation provided by EM facilities (discovery time, $t_{\rm disc}$ ) is used as the time of SBO and the end of the OSW, $t_2$. The beginning of the OSW is calculated from the last EM null observation of a CCSN in the host galaxy, $t_{\rm null}$. However, we should not use $t_{\rm null}$ as $t_1$, because even if it is before SBO, it may be after core bounce and the initiation of the GW emission, which in turn would have left the exploding progenitor before SBO and arrived at the Earth before $t_{\rm null}$ (it is also important to asses that the sensitivity of the available telescope observations is not impairing the capability to ensure that $t_{\rm null}$ is really before the SBO \cite{Gill:2022amm}). $t_{1}$ can be chosen to be $t_{\rm null}$ minus the delay between collapse and SBO.  An upper limit for the delay between collapse and SBO could be inferred from \citet{Barker:2021iyr} given the progenitor-mass estimate. If someone is sufficiently confident in the progenitor-mass estimate and mass--delay conversion, $t_2$ can be moved closer to $t_1$ by using the lower limit in the delay from \citet{Barker:2021iyr}. Overall the decision is a trade off between making sure that the OSW contains the GW emission and maximizing the detection capabilities.

(ii) Quartic Interpolation:
If the light curve is discovered in the homologous phase of the shock expansion we can use the fact that it has in this phase a high degree of universality among observed CCSNe. In \citet{Szczepanczyk:2023ihe} the use of the simplest polynomial fit was investigated (where the coefficients of different polynomials were estimated with a chi-squared minimization between the data and the polynomials). Such a fit was used, as a testing ground, on a detailed light curve (from the moment of observed SBO) obtained for KSN2011a by Kepler \cite{Garnavich:2016thk} and Astropy (\citet{Astropy:2013muo,Astropy:2018wqo,Astropy:2022ucr}). Figure 3 of \citet{Szczepanczyk:2023ihe} shows possible interpolations in cases where a potential first observation can be up to a few days after SBO. A simple quadratic polynomial fit revealed biases. In this case, the estimated time of SBO was earlier than the actual time of SBO because the slope of the light curve around its peak is smaller than in its early stages. A quartic polynomial fit is instead able to more accurately estimate SBO even when the first few days of data are missing.

(iii) Physics-Based:
Another approach used for some of the candidates in \citet{Szczepanczyk:2023ihe} and described in more detail in \citet{Gill:2022amm} is to estimate the time of SBO with light curves based on either physical modelling or matched filtering with template banks of recorded light curves. This approach also provides posterior distributions for the possible time of arrival of the GW at Earth.

For targeted CCSN searches with neutrino triggers versus EM triggers, the data associated with the seconds-long OSWs will be available for the whole time interval, and it is straightforward to choose an OSW that is both short and has a $100\%$ probability of containing the GW emission. On the other hand if EM triggers are used (as well as for all-sky searches) a sizable fraction of the days-long OSW will not be available in at least one of the interferometers. The so-called duty cycle of available science-quality coincident data can vary from $20\%$ to close to $90\%$ (see examples in 
\citet{Szczepanczyk:2023ihe}). Furthermore it is currently hard to quantify precisely all the uncertainties associated with the OSW from an EM observation. Therefore for an EM-targeted search, having an OSW that has a probability between $90\%$ and $100\%$ to contain the GW emission and increase the detection chances is preferable to a very long OSW with no detection benefit, and in fact such a strategy was adopted in the most recent search \citet{Szczepanczyk:2023ihe} (also because the data outside the OSW is processed in parallel, with lower sensitivity, by the all-sky searches). Finally, the OSW discussion is also relevant for all-sky searches in order to evaluate the possible correlation of candidate events with known CCSNe.

\vspace{-10pt}
\section{Conclusions and Outlook}
\label{sec:ConclusionsandOutlook}

In this review, we hope we have conveyed the status of CCSN modeling and GW prediction, efforts to detect a Galactic or near-extra-Galactic event, and efforts to infer defining characteristics of the progenitor and supernova central engine given a detection. Three-dimensional models have reached a significant level of sophistication, though important work remains, especially to elevate all of the leading models across groups to full three-dimensionality and full general relativity. Supercomputing architectures are enabling these advances. Our understanding of CCSN GW emission has grown dramatically in recent years, in sync with the increase in the sophistication of explosion modeling and, most important, full physics three-dimensional modeling. Lessons from GW detection and PE of CBCs have provided a foundation on which detection and PE methods specifically designed for CCSNe have been developed. The advances in modeling mentioned above have aided this development, providing significant incite into the GW features associated with stellar core bounce, PNS convection, neutrino-driven convection, the SASI, and explosion. Fortunately, in some cases, the expected spectral characteristics of GW emission from these central engine components lie within the most sensitive parts of the interferometer sensitivity curves. And the development of GW data analysis has helped hone CCSN simulation, as well, in order to provide the targeted, optimal output for GW detection and PE.

GW astronomy is just beginning and, even more so, CCSN GW astronomy. The challenges we face, in light of the weakness of the GW emission from CCSNe and the paucity of such events in the Galaxy, are significant. But the scientific payload is too great to withdraw from such challenges. Thus far, the community has not, and we hope we have conveyed the significant progress that has been made along all three necessary fronts: modeling, detection, and PE. Given this progress, and given that we may be overdue for a Galactic event, these are truly exciting times to be modeling CCSNe and their GW emission, and to be investing in methods, tools, and instruments designed to see them in ways only GWs can enable, to the very center of the supernova central engine.

\vspace{-10pt}
\section{Acknowledgments}
A.M. was supported in part by the National Science Foundation Gravitational Physics Theory Program through awards PHY 1806692 and PHY 2110177. M.Z. was supported by the National Science Foundation Gravitational Physics Experimental and Data Analysis Program through award PHY 2110555. The authors would like to express their thanks to Haakon Andresen, Viktoriya Morozova, and David Radice for kindly letting us include their figures in this review.

\end{document}